\documentclass[11pt,english, mathscr,psamsfonts,cmex10,makeidx]{article}
\usepackage{helvet}
\usepackage[latin1]{inputenc}
\usepackage{geometry}
\usepackage{amsmath}
\usepackage{setspace}
\usepackage{amssymb}
\usepackage[authoryear]{natbib}

\makeatletter

\providecommand{\tabularnewline}{\\}

\usepackage{sublabel}
\usepackage{mathpi}
\usepackage{amssymb}
\usepackage{geometry}
\allowdisplaybreaks[1]
\usepackage{setspace}
\usepackage{amsfonts}
\usepackage{latexsym}
\usepackage{amsthm}
\usepackage{graphicx}
\usepackage{multirow}
\usepackage{color}
\usepackage{subfigure}
\usepackage{index}
\usepackage{fancybox}
\usepackage{oldgerm}
\usepackage{euscript}
\usepackage{exscale}
\usepackage{pifont}
\usepackage{showidx}
\usepackage{array}
\usepackage{afterpage}
\usepackage{bm}
\makeindex

\AtBeginDocument{
  
}

\makeatother
\begin{document}
{\tiny .}{\tiny \par}
\vspace{1cm}

\noindent \begin{center}\textbf{\Huge ChaNoXity: The Nonlinear Dynamics of Nature}\end{center}{\Huge \par}

\noindent \begin{center}A. Sengupta\\Department of Mechanical Engineering\\Indian
Institute of Technology Kanpur, Kanpur 208016, INDIA.\\E-Mail: osegu@iitk.ac.in\end{center}

\begin{abstract}
\noindent In this paper we employ the topological-multifunctional mathematical
lanNguage and techniques of non-injective illposedness developed in \citet{Sengupta2003}
to formulate a notion of \emph{ChaNoXity} --- Chaos-Nonlinearity-Complexity
--- in describing the specifically nonlinear dynamical evolutionary processes
of Nature. Non-bijective ill-posedness is the natural mode of expression for
chanoxity that aims to focus on the nonlinear interactionsd generating dynamical
evolution of real irreversible processes. The basic dynamics is considered to
take place in a matter-antimatter \emph{kitchen space} $X\times\mathfrak{{X}}$
of Nature that is inaccessible to both the functional matter $(X)$ and multifunctional
antimatter $(\mathfrak{{X}})$ components. These component spaces are distinguished
by opposing evolutionary directional arrows and satisfy the defining property\[
(\forall A\subseteq X,\,\exists\,\mathfrak{{A}}\subseteq\mathfrak{{X}})\,\textrm{s.t.}\,(A\cup\mathfrak{{A}}=\emptyset).\]
Dynamical equilibrium is considered to be represented by such \emph{competitively
collaborating} stasis states of the matter-antimatter constituents of Nature.
\end{abstract}

\bigskip{}
\section{\textsf{\label{sec: 1. Introduction}}Introduction}

This paper applies the mathematical language and techniques of non-bijective,
and in particular non-injective, ill-posedness developed in \citet{Sengupta2003}
to formulate an integrated approach to chaos, nonlinearity and complexity (ChaNoXity),
where a complex system is taken to be characterized by 

\smallskip{}
$\blacktriangleright$ a collection of many \emph{interdependent} \emph{parts }

$\blacktriangleright$ that interact with each other through \emph{competitive}
\emph{nonlinear} \emph{collaboration} 

$\blacktriangleright$ leading to \emph{emergent}, \emph{self}-\emph{organized}
behaviour.%
\footnote{Competitive collaboration --- as opposed to reductionism --- in the context
of this characterization is to be understood as follows: The interdependent
parts retain their individual identities, with each contributing to the whole
in its own characteristic fashion \emph{within the framework of global properties
of the union}. \emph{}A comparison of reductionism as summarized in Figs. \ref{fig: cmplx_2cycle},
\emph{b,} and \emph{c}, shows that although the properties of the whole are
generated by the parts, these units acting independently on their own cannot
account for the emergent global behaviour of the whole. %
}
\smallskip{}

\noindent We will show how each of these defining characteristics of complexity
can be described and structured within the mathematical framework of our multifunctional
graphical convergence of a net of functions $(f)_{\alpha}$. In this programme,
convergence in topological spaces continues to be our principal tool, and the
particular topologies of significance that emerge are the topology of saturated
sets and the $A$-exclusion topology, with $A$ a subset of the domain of $f_{\alpha}$.
We will demonstrate that a complex system can be described as an association
of independent expert groups, each entrusted with a specific specialized task
by a top-level central coordinating command, that consolidates and regulates
the inputs received from its different constituent units by harmonizing and
combining them into an emerging whole; thus the complexity of a system, broadly
speaking, is the amount of information needed to describe it. In this task,
and depending on the evolving complexity of the dynamics, the central unit delegates
its authority to subordinate units that report back to it the data collected
at its own level of authority. 

Recall that (i) a multifunction --- which constitutes one of the foundational
notions of our work --- and the non-injective function are connected by \begin{align}
f\textrm{ is a non-injective function} & \Longleftrightarrow f^{-}\textrm{ is a multifunction}\label{eq: func-multifunc}\\
f\textrm{ is a multifunction} & \Longleftrightarrow f^{-}\textrm{ is a non-injective function}.\nonumber \end{align}
and (ii) the neighbourhood \emph{}of a point $x\in(X,\mathcal{U})$ --- which
is a generalization of the familiar notion of distances of metric spaces ---
is a nonempty subset $N$ of $X$ containing an open set $U\in\mathcal{U}$;
thus $N\subseteq X$ is a neighbourhood of $x$ iff $x\in U\subseteq N\subseteq(X,\mathcal{U})$
for some open set $U$ of $X$. The collection of all neighbourhoods of $x$
\begin{equation}
\mathcal{N}_{x}\overset{\textrm{def}}=\{ N\subseteq X\!:x\in U\subseteq N\textrm{ for some }U\in\mathcal{U}\}\label{eq: Def: nbd system}\end{equation}
 \emph{}is the \emph{}neighbourhood system at $x$, and the subcollection $U$
of $\mathcal{U}$ used in this expression constitutes a \emph{}neighbourhood
(local) base \emph{}or \emph{}basic neighbourhood system, at \emph{}$x$. The
properties 

\smallskip{}
(N1) $x$ belongs to every member $N$ of $\mathcal{N}_{x}$\emph{, }

(N2) The intersection of any two neighbourhoods of $x$ \emph{}is another neighbourhood
of $x$: $N,M\in\mathcal{N}_{x}\Rightarrow N\cap M\in\mathcal{N}_{x}$, 

(N3) Every superset of \emph{}any neighbourhood of $x$ is a neighbourhood of
$x$: $(M\in\mathcal{N}_{x})\wedge(M\subseteq N)\Rightarrow N\in\mathcal{N}_{x}$ 

\smallskip{}
\noindent characterize $\mathcal{N}_{x}$ \emph{}completely and imply that a
subset \textsl{$G\subseteq(X,\mathcal{U})$} is open iff it is a neighbourhood
of each of its points. Accordingly if $\mathcal{N}_{x}$ is an arbitrary collection
of subsets of $X$ associated with each $x\in X$ satisfying $(\textrm{N}1)-(\textrm{N}3)$,
then the special class of neighbourhoods $G$ \begin{equation}
\mathcal{U}=\{ G\in\mathcal{N}_{x}\!:x\in B\subseteq G\textrm{ for some }B\in\mathcal{N}_{x}\textrm{ and each }x\in G\}\label{eq: nbd-topology}\end{equation}
 defines a unique topology on $X$ containing a basic neighbourhood $B$ at
each of its points $x$ for which the neighbourhood system \textit{\emph{is
the prescribed collection}} $\mathcal{N}_{x}$\textit{.} Among the three properties
$(\textrm{N}1)-(\textrm{N}3)$, the first two now re-expressed as 

\smallskip{}
(NB1) $x$ belongs to each member $B$ of $\mathcal{B}_{x}$\emph{. }

(NB2) The intersection of any two members of \emph{}$\mathcal{B}_{x}$ \emph{}contains
another member of $\mathcal{B}_{x}$: $B_{1},B_{2}\in\mathcal{B}_{x}\Rightarrow(\exists B\in\mathcal{B}_{x}\!:B\subseteq B_{1}\cap B_{2})$. \emph{}
\smallskip{}

\noindent are fundamental in the sense that the resulting subcollection $\mathcal{B}_{x}$
of $\mathcal{N}_{x}$ generates the full system by appealing to $(\textrm{N}3)$.
This \emph{}basic neighbourhood \emph{}system, or \emph{}local base, \textsl{}at
\textsl{$x$} in \textsl{$(X,\mathcal{U})$} satisfies \begin{align}
\mathcal{B}_{x} & \overset{\textrm{def}}=\{ B\in\mathcal{N}_{x}\!:x\in B\subseteq N\textrm{ for each }N\in\mathcal{N}_{x}\}\label{eq: TBx}\end{align}
 \textsl{}which reciprocally determines the full neighbourhood system \textsl{}\begin{equation}
\mathcal{N}_{x}=\{ N\subseteq X\!:x\in B\subseteq N\textrm{ for some }B\textrm{ }\in\,\mathcal{B}_{x}\}\label{eq: TBx_nbd}\end{equation}

\noindent as all the supersets of these basic elements.

The topology of saturated sets is defined in terms of equivalence classes $[x]_{\sim}=\{ y\in X\!:y\sim x\in X\}$
generated by a relation $\sim$ on a set $X$; the neighbourhood system $\mathcal{N}_{x}$
of $x$ in this topology consists of all supersets of the equivalence class
$[x]_{\sim}\in X/\sim$. In the $x$-exclusion topology of all subsets of $X$
that exclude $x$ (plus $X$, of course), the neighbourhood system of $x$ is
just $\{ X\}$. While the first topology provides, as in \citet{Sengupta2003},
the motive force for an evolutionary direction in time, the second will define
an \emph{anti-space} $\mathfrak{{X}}$ of (associated with; generated by) $X$,
with an oppositely directed evolutionary arrow. With \emph{dynamic equilibrium}
representing a state of stasis between the associated opposing motives of evolution,
\emph{(static) equilibrium} will be taken to mark the end of a directional evolutionary
process represented by convergence of the associated sequence to an adherence
set. 

Let $f\!:X\rightarrow Y$ be a function and $f^{-}\!:Y\quad\; X$ its multi-inverse:
hence $ff^{-}f=f$ and $f^{-}ff^{-}=f^{-}$ although $f^{-}f\neq\mathbf{1}_{X}$
and $ff^{-}\neq\mathbf{{1}}_{Y}$ necessarily. Some useful identities for subsets
$A\subseteq X$ and $B\subseteq Y$ are shown in Table \ref{tab: ill-posed},
where the complement of a subset $A\subseteq X$ is denoted by $A^{c}=\{ x\!:(x\in X-A)\wedge(x\not\in A)\}$.
Let the \emph{$\textrm{f}$-saturation} $\mathscr{S}_{f}(A):=f^{-}f(A)$ of
$A$ and \emph{}the \emph{$\textrm{f}$-component} $\mathscr{C}_{f}(B):=ff^{-}(B)=B\cap f(X)$
of $B$ on the image of $f$ define generalizations of injective and surjective
mappings in the sense that any $f$ behaves one-one and onto on its saturated
and component sets respectively, so that it possible to replace each of the
relevant assertions of Table \ref{tab: ill-posed} with the more direct injectivity
and surjectivity conditions on $f$. Indeed\begin{eqnarray*}
f(x)=y & \Longrightarrow & f(f^{-}f(x))=y=ff^{-}(y)\\
 & \Longrightarrow & f(\mathscr{S}_{f}(x))=\mathscr{C}_{f}(y)\end{eqnarray*}
and \begin{eqnarray*}
x=f^{-}(y) & \Longrightarrow & f^{-}f(x)=x=f^{-}(ff^{-}(y))\\
 & \Longrightarrow & \mathscr{S}_{f}(x)=f^{-}(\mathscr{C}_{f}(y))\end{eqnarray*}
 proves the bijectivity of $f\!:\mathscr{S}_{f}(x)\rightarrow\mathscr{C}_{f}(y)$
restricted to $\mathscr{S}_{f}(x)$ and $\mathscr{C}_{f}(y)$; hence in the
bijective inverse notation the corresponding functional equation takes the form

\begin{equation}
f(\mathscr{S}_{f}(A))=\mathscr{C}_{f}(B)\Longleftrightarrow\mathscr{S}_{f}(A)=f^{-1}(\mathscr{C}_{f}(B)).\label{eq: bijective_homeo}\end{equation}
These important generalizations of the bijectivity of functions are of great
value to us because our notion of chaos and complexity is based on ill-posedness
of the non-bijective type of functional equations $f(x)=y$. %
\begin{table}[htbp]
\noindent {\renewcommand{\arraystretch}{1.3}

\noindent \begin{center}\begin{tabular}{|r|c|c|}
\hline 
&
$f\!:X\rightarrow Y$&
$f^{-}\!:Y\quad X$\tabularnewline
\hline
\hline 
1&
$A_{1}\subseteq A_{2}\Rightarrow f(A_{1})\subseteq f(A_{2})$&
$B_{1}\subseteq B_{2}\Rightarrow f^{-}(B_{1})\subseteq f^{-}(B_{2})$\tabularnewline
\multicolumn{1}{|r|}{}&
$\qquad\quad\;\,\Leftarrow\textrm{ iff }A=\mathscr{S}_{f}(A)$&
$\qquad\;\Leftarrow\textrm{ iff }B=\mathscr{C}_{f}(B)$\tabularnewline
\hline 
\multicolumn{1}{|r|}{2}&
$f(A)\subseteq B\Longleftrightarrow A\subseteq f^{-}(B)$&
$f(A)\subseteq B\Longleftrightarrow A\subseteq f^{-}(B)$\tabularnewline
\multicolumn{1}{|r|}{}&
$B\subseteq f(A)\Rightarrow f^{-}(B)\subseteq A\textrm{ iff }A=\mathscr{S}_{f}(A)$&
$B\subseteq f(A)\Leftarrow f^{-}(B)\subseteq A\textrm{ iff }B=\mathscr{C}_{f}(B)$\tabularnewline
\hline
3&
$A=\emptyset\Leftrightarrow f(A)=\emptyset$&
$f^{-}(\emptyset)=\emptyset$\tabularnewline
&
&
$f^{-}(B)=\emptyset\Rightarrow B=\emptyset$ iff $B=\mathscr{C}_{f}(B)$\tabularnewline
\hline
4&
$f(A_{1})\cap f(A_{2})=\emptyset\Rightarrow A_{1}\cap A_{2}=\emptyset$&
$f^{-}(B_{1})\cap f^{-}(B_{2})=\emptyset\Leftarrow B_{1}\cap B_{2}=\emptyset$\tabularnewline
&
$\qquad\qquad\qquad\;\;\qquad$$\Leftarrow$ iff $A=\mathscr{S}_{f}(A)$&
$\qquad\qquad\qquad\qquad\qquad\;\;$$\Rightarrow$ iff $B=\mathscr{C}_{f}(B)$\tabularnewline
\hline
5&
$f(\cup_{\alpha}A_{\alpha})=\cup_{\alpha}f(A_{\alpha})$&
$f^{-}(\cup_{\alpha}B_{\alpha})=\cup_{\alpha}f^{-}(B_{\alpha})$\tabularnewline
6&
$f(\cap_{\alpha}A_{\alpha})\subseteq\cap_{\alpha}f(A_{\alpha}),$ ''='' iff
$A=\mathscr{S}_{f}(A)$&
 $f^{-}(\cap_{\alpha}B_{\alpha})=\cap_{\alpha}f^{-}(B_{\alpha})$ \tabularnewline
7&
$f(A^{c})=(f(A))^{c}\cap f(X)$ iff $A=\mathscr{S}_{f}(A)$&
$f^{-}(B^{c})=((f^{-}(B))^{c}$\tabularnewline
\hline
\end{tabular}\end{center}

\caption{{\small \label{tab: ill-posed}The role of saturated and component sets in
a function and its inverse; here all} {\footnotesize }{\small $A=\mathscr{S}_{f}(A)$
and $B=\mathscr{C}_{f}(B)$ are to be understood to hold for every $A\subseteq X$
and $B\subseteq Y$. Unlike $f$, $f^{-}$ preserves the basic set operations
in the sense of 5, 6, and 7. This makes $f^{-}$ rather than $f$ the ideal
instrument for describing topological and measure theoretic properties like
continuity and measurability of functions.}}}
\end{table}

All statements of the first column of the table for saturated sets $A=\mathscr{S}_{f}(A)$
apply to the quotient map $q$; observe that $q(A^{c})=(q(A))^{c}$. Moreover
combining the respective entries of both the columns, it is easy to verify the
following results for the saturation map $\mathscr{S}_{f}=f$ on saturated sets
$A=\mathscr{S}_{f}(A)$. 

\smallskip{}
(a) $\mathscr{S}_{f}\left({\textstyle \bigcup}\, A_{i}\right)={\textstyle \bigcup}\,\mathcal{S}_{f}(A_{i})$:
The union of saturated sets is saturated. 

(b) $\mathscr{S}_{f}\left({\textstyle \bigcap}\, A_{i}\right)={\textstyle \bigcap}\,\mathcal{S}_{f}(A_{i})$:
The intersection of saturated sets is saturated. 

(c) $X-\mathscr{S}_{f}(A)=\mathcal{S}_{f}(X-A)$: The complement of a saturated
set is saturated. 

(d) $A_{1}\subseteq A_{2}\Rightarrow\mathscr{S}_{f}(A_{1})\subseteq\mathscr{S}_{f}(A_{2})$

(e) $\mathscr{S}_{f}\left({\textstyle \bigcap}\, A_{i}\right)=\emptyset\Rightarrow{\textstyle \bigcap}\, A_{i}=\emptyset.$
\smallskip{}

While properties (a) and (b) lead to the topology of saturated sets, the third
makes it a complemented topology when the (closed) complement of an open set
is also an open set. In this topology there are no boundaries between sets which
are isolated in as far as a sequence eventually in one of them converging to
points in the other is concerned. 

Since the guiding incentive for this work is an understanding of the precise
role of irreversibility and nonlinearity in the dynamical evolution of irreversible
real processes, we will propose an index of nonlinear irreversibility in essentially
the kitchen \emph{}space $X\times\mathfrak{{X}}$ of Nature, wherein all the
evolutionary dynamics are postulated to take place. The real world $X$ is only
a projection of this multifaceted kitchen that is distinguished in having a
non-real anti-component $\mathfrak{{X}}$ interacting with $X$ to generate
the dynamical reality perceived in the later. This nonlinearity index, together
with the dynamical stasis between opposing directional arrows associated with
$X$ and its anti-world $\mathfrak{{X}}$, suggests a description of time's
arrow that, unlike both the historical and modern entropic approaches, is specifically
nonlinear with chaos and complexity being the prime manifestations of strongly
nonlinear systems. 

The entropy produced within a system due to irreversibilities within it \citep{Kondepudi1998}
are generated by nonlinear dynamical interactions between the system and its
anti-world, and the objective of this paper is to clearly define this interaction
and focus on its relevance in the dynamical evolution of  Nature.

\section{\label{sec: Chanoxity}ChaNoXity: Chaos-Nonlinearity-Complexity}

\subsection{\label{sub: Entr-Irrev-Non}Entropy, Irreversibility, and Nonlinearity}

In this subsection we summarize the {}``modern'' approach to entropy due to
De Donder as enunciated by \citet{Kondepudi1998} which explicitly incorporates
irreversibility into the formalism of the Second Law of Thermodynamics thereby
making it unnecessary to consider ideal, non-real, reversible processes for
computing (changes in) entropy. This follows from the original Clausius inequality
\[
dS\geq\frac{dQ}{T}\]
that may be written in the form \begin{equation}
dS=\frac{dQ}{T}+dI\label{eq: uncompensated}\end{equation}
where $dQ/T$ is due to the heat exchanged by the system with its exterior and
$I$, the {}``uncompensated transformation'' of Clausius, represents the entropy
produced from the real irreversible processes occurring within the system. In
postulating the existence of an entropy function $S(U,V,N)$ of the extensive
parameters of internal energy $U$, volume $V$, and mole numbers $\{ N\}_{j=1}^{J}$
of the chemical constituents comprising a composite system that is defined for
all equilibrium states, we follow \citet{Callen1985} in supposing that in the
absence of internal constraints the extensive parameters assume such values
that maximize $S$ over all the constrained equilibrium states. The entropy
of the composite system is additive over the constituent subsystems, and is
continuous, differentiable, and increases monotonically with respect to the
energy $U$. This last property implies that $S(U,V,N)$ can be inverted in
$U(S,V,N)$; hence \begin{equation}
dU(S,V,\{ N_{j}\})=\frac{\partial U}{\partial S}\, dS+\frac{\partial U}{\partial V}\, dV+\sum_{j=1}^{J}\frac{\partial U}{\partial N_{j}}\, dN_{j}\label{eq: dU_differential}\end{equation}
defines the \emph{intensive parameters}\sublabon{equation}\begin{eqnarray}
{\displaystyle \frac{\partial U}{\partial S}} & \!\!\overset{\textrm{def}}=\!\! & T(S,V,\{ N_{j}\}_{j=1}^{J}),\qquad V,\,\{ N_{j}\}\,\textrm{held const}\label{eq: T-P-mu_def (a)}\\
{\displaystyle \frac{\partial U}{\partial V}} & \!\!\overset{\textrm{def}}=\!\! & -P(S,V,\{ N_{j}\}_{j=1}^{J}),\qquad S,\,\{ N_{j}\}\,\textrm{held const}\label{eq: T-P-mu_def (b)}\\
{\displaystyle \frac{\partial U}{\partial N_{j}}} & \!\!\overset{\textrm{def}}=\!\! & \mu_{j}(S,V,\{ N_{j}\}_{j=1}^{J}),\qquad S,\, V\,\textrm{held const}\label{eq: T-P-mu_def (c)}\end{eqnarray}
\sublaboff{equation}of absolute temperature $T$, pressure $P$, and chemical
\emph{}potential $\mu_{j}$ of the $j^{\textrm{th}}$ component, from the macroscopic
extensive ones. Inversion of Eq. (\ref{eq: dU_differential}) gives the differential
\emph{Gibbs entropy} definition \begin{eqnarray}
dS(U,V,\{ N_{j}\}) & \!\!\overset{\textrm{def}}=\!\! & {\displaystyle \frac{1}{T(U,V,\{ N_{j}\})}}\, dU+{\displaystyle \frac{P(U,V,\{ N_{j}\})}{T(U,V,\{ N_{j}\})}}\, dV-{\displaystyle {\displaystyle \sum_{j=1}^{J}\frac{\mu_{j}(U,V,\{ N_{j}\})}{T(U,V,\{ N_{j}\})}\, dN_{j}}}\label{eq: Gibbs_Entropy}\end{eqnarray}

\noindent providing an equivalent correspondence of the partial derivatives
$(\partial S/\partial U)_{V,N_{j}}=1/T(U,V,\{ N_{j}\})$, $(\partial S/\partial V)_{U,N_{j}}=P(U,V,\{ N_{j}\})/T$,
and $(\partial S/\partial N_{j})_{U,V}=-\sum_{j=1}^{J}\mu(U,V,\{ N_{j}\})/T$
with the intensive variables of the system. 

In the spirit of the Pffafian differential form, dependence of the intensive
variables of the First Law \begin{eqnarray}
dU(S,V,\{ N_{j}\}) & \!\!=\!\! & dQ(S,V,\{ N_{j}\})+dW(S,V,\{ N_{j}\})+dM(S,V,\{ N_{j}\}),\nonumber \\
 & \!\!=\!\! & dQ(S,V,\{ N_{j}\})-P(S,V,\{ N_{j}\})\, dV+{\displaystyle \sum_{j=1}^{J}\mu_{j}(S,V,\{ N_{j}\})\, dN_{j}}\label{eq: 1-Law}\end{eqnarray}
 --- that can be taken to define the heat flux $dQ$ --- on the respective extensive
macroscopic variables $U$, $V$, or $N_{j}$ serves to decouple the (possibly
nonlinear) bonds between them; this is necessary and sufficient for the resultant
thermodynamics to be classified as \emph{quasi-static} or \emph{reversible}.
\emph{}These ideal states as pointed out by \citet{Callen1985} are simply an
ordered class of equilibrium states, neutral with respect to time-reversal and
without any specific directional properties, that is distinguished from natural
real processes of ordered \emph{temporal} \emph{successions} of equilibrium
and non-equilibrium \emph{}states: \emph{a reversible quasi-static process is
simply a directionless collection of elements of an ordered set.}%
\footnote{\label{foot: Uffink}Jos Uffink \citeyearpar{Uffink2001} delineates three different
types of (ir)reversibilities. The most comprehensive among these follows from
the notion of a \emph{time-(a)symmetric theory} that requires the (non)existence
of a reverse process $\mathscr{P}_{r}:=\{ r(-t)\!:-t_{f}\leq t\leq-t_{i}\}$
for every permissible forward process $\mathscr{P}:=\{ s(t)\!:t_{i}\leq t\leq t_{f}\}$
of the theory; here $r=Rs$ with $R^{2}=\mathbf{1}$, is the time-reversal of
state $s$. Although in contrast with mechanics thermodynamics has no equations
of motion, the Second Law endows it with a time-asymmetric character and a thermodynamic
process is irreversible iff its reverse $\mathscr{P}_{r}$ is not allowed by
the theory. 

Two weaker concepts of reversibility --- requiring only that the system and
its environment be restored to the respective initial conditions by the reverse
process without any reference to the intermediate states, and quasi-staticity
in which the process is poised so delicately as to proceed infinitely slowly,
effectively in equilibrium throughout --- are more common in thermodynamics.
Our use of the notion of irreversibility in Sec. \ref{sub: Antispace} will
be in the spirit of time-irreversibility. %
} From the definition Eq. (\ref{eq: T-P-mu_def (a)}) of the absolute temperature
$T$, it follows that \emph{under quasi-static conditions}\begin{equation}
dQ(S)\overset{\textrm{def}}=T(S)\, dS,\label{eq: dS_rev}\end{equation}

\noindent reduces the heat transfer $dQ$ to formally behave work-like that
permits Eq. (\ref{eq: 1-Law}) to be expressed in the \emph{combined first and
second law} form \sublabon {equation}\begin{eqnarray}
dU(S,V,\{ N_{j}\}) & \!\!=\!\! & T(S)\, dS-P(V)\, dV+\sum_{j=1}^{J}\mu(N_{j})\, dN_{j}\label{eq: dU_quasi-static(a)}\\
dS(U,V,\{ N_{j}\}) & \!\!=\!\! & \frac{1}{T(U)}\, dU+\frac{P(V)}{T(U)}\, dV-\sum_{j=1}^{J}\frac{\mu(N_{j})}{T(U)}\, dN_{j}\label{eq: dU_quasi-static(b)}\end{eqnarray}
\sublaboff {equation}which are just the integrable \emph{quasi-static versions}
of Eqs. (\ref{eq: dU_differential}, \ref{eq: Gibbs_Entropy}). Note that the
total energy input and the corresponding entropy transfer in the quasi-static
case reduces to a simple sum of the constituent parts of the change. For non
quasi-static real processes, this linear superposition of the solution into
its individual components is not justified as the solution of the resulting
Pfaffian equation is the general $U(S,V,\{ N_{j}\}_{j}=\textrm{const}$. For
any natural non-cyclic real process therefore, the identification \begin{equation}
dQ(S,V,\{ N_{j}\})\overset{\textrm{def}}=T(S,V,\{ N_{j}\})\, dS\label{eq: dS_irrev}\end{equation}
reduces (\ref{eq: dU_differential}) to the first law form (\ref{eq: 1-Law})
for real processes that no longer be decomposes into individual and non-interacting
heat, mechanical work, and mass transforming processes of its quasi-static counterpart
(\ref{eq: dU_quasi-static(a)}). Eq. (\ref{eq: dS_irrev}) is graphically expressed
\citep{Kondepudi1998} in the spirit of (\ref{eq: uncompensated}) as\begin{align}
dS & =\frac{dQ(S,V,\{ N_{j}\})}{T(S,V,\{ N_{j}\})}\nonumber \\
 & =\frac{d\widehat{Q}}{T}+\frac{d\widetilde{Q}}{T}\nonumber \\
 & =d\widehat{S}+d\widetilde{S},\label{eq: dS eq d{_e}S+d{_i}S}\end{align}
where the total entropy exchange is expressed as a sum of two parts: the first
\[
d\widehat{S}=\frac{d\widehat{Q}}{T}\gtrless0\]
 may be positive, zero or negative depending on the specific nature of energy
transfer $d\widehat{Q}$ with the (infinite) exterior reservoir, but the second
\begin{equation}
d\widetilde{S}=\frac{d\widetilde{Q}}{T}\geq0\label{eq: entropy_internal}\end{equation}
 representing the entropy produced by irreversible nonlinear processes within
the system is always positive. Expressing $dQ$ by the first law Eq. (\ref{eq: 1-Law})
in terms of the basic macroscopic extensive variables $U$, $V$, and $N$,
yields for a composite body $C=A\cup B$ of two parts $A$ and $B$, each interacting
with its own infinite reservoir under the constraint $U=U_{A}+U_{B}$, $V=V_{A}+V_{B}$
and $N=N_{A}+N_{B}$, the Gibbs expression \begin{align}
dS_{C}(U,V,N) & =\left[\frac{1}{T_{A}}\, dU_{A}+\frac{1}{T_{B}}\, dU_{B}\right]+\left[\frac{p_{A}}{T_{A}}\, dV_{A}+\frac{p_{B}}{T_{B}}\, dV_{B}\right]-\left[\frac{\mu_{A}}{T_{A}}\, dN_{A}+\frac{\mu_{B}}{T_{B}}\, dN_{B}\right]\label{eq: d{e_}S}\end{align}
for the entropy exchanged by $C$ in reaching a state of \emph{static} \emph{equilibrium}
with its infinite environment; here $T$, $P$ and $\mu$ are the parameters
of the reservoirs that completely determine the internal state of $C$. This
exchange of energy with the surroundings perturbs the system from its state
of equilibrium and sets up internal irreversible nonlinear processes between
the two subsystems, driving $C$ towards a new state of \emph{dynamic} \emph{equilibrium}
that can be represented (\citet{Katchalsky1965}, \citet{Kondepudi1998}) in
terms of flows of extensive quantities set up by forces generated by the intensive
variables. Thus for a composite \emph{dynamically} interacting system $C=A\cup B$
consisting, for example, of two chambers $A$ and $B$ of volumes $V_{A}$ and
$V_{B}$ containing two nonidentical gases at distinct temperatures, pressures,
and mole numbers, the entropy generated by nonlinear irreversible processes
within the system when the partition separating the chambers is removed, can
be expressed in the Gibbs form as \begin{align}
d\widetilde{S_{C}}(U,V,N) & =\left[\frac{1}{T_{A}}-\frac{1}{T_{B}}\right]dU_{A}+\left[\frac{p_{A}}{T_{A}}-\frac{p_{B}}{T_{B}}\right]dV_{A}-\left[\frac{\mu_{A}}{T_{A}}-\frac{\mu_{B}}{T_{B}}\right]dN_{A},\label{eq: d{i_}S}\\
 & U=U_{A}+U_{B}=\textrm{const},\, V=V_{A}+V_{B}=\textrm{const},\, N=N_{A}+N_{B}=\textrm{const}\nonumber \end{align}
with each term on the right, a product of an intensive thermodynamic force driving
the corresponding extensive thermodynamic flow, contributing to the \emph{uncompensated
heat} of Clausius, see \citet{Kondepudi1998}. This uncompensated heat generated
entirely within the system due to the nonlinear irreversible dynamical interactions
between $A$ and $B$ is taken to be responsible for the increase of entropy
accompanying all natural processes. This interaction between two (finite) systems
is to be compared and contrasted with the static interaction between a (finite)
system and an (infinite) reservoir. Compared with the later for which the time
evolution is unidirectional with the system unreservedly acquiring the properties
of the reservoir that undergoes no perceptible changes leading to a state of
\emph{static equilibrium} as a result of the \emph{passive interaction} of the
system with its reservoir, the system-system interaction is fundamentally different
as it evolves bidirectionally such that the properties of the composite are
not of either of the subsystems, but an average of the individual properties
defining an eventual state of \emph{dynamic interactive equilibrium}. This distinction
between passive and dynamical interactive equilibria resulting respectively
from the uni- and bi-directional interactions is clearly revealed in Eqs. (\ref{eq: d{e_}S})
and (\ref{eq: d{i_}S}), with bi-directionality of the later being displayed
by the \emph{difference form of the generalized forces}. Accordingly, subsystem
$A$ (respectively subsystem $B$) has two directional arrows imposed on it:
the first from its own forward evolution that is opposed by a second reverse
process due to its interactive interaction with $B$ (respectively $A$), see
Fig. \ref{fig: stasis}. Evolution requires all macroscopic extensive variables
--- and hence all the related microscopic intensive parameters --- to be functions
of time so that equilibrium, in the case of Eq. (\ref{eq: d{i_}S}) for example,
demands \begin{align}
\frac{d\widetilde{S_{C}}}{dt}=0 & \Longrightarrow\left(\frac{dU_{A}}{dt}=0\right){\textstyle {\textstyle \bigwedge}}\left(\frac{dV_{A}}{dt}=0\right){\textstyle {\textstyle \bigwedge}}\left(\frac{dN_{A}}{dt}=0\right)\nonumber \\
 & \Longleftarrow(T_{A}(t)=T_{B}(t))\wedge(p_{A}(t)=p_{B}(t))\wedge(\mu_{A}(t)=\mu_{B}(t)).\label{eq: stasis}\end{align}

While we return to this topic subsequently using the tools of directed sets
and convergence in topological spaces, for the present it suffices to note that
for an emerging, complex, self-organizing, evolving system of the type that
concerns us here, the linear reductionist decoupling of its entropy change into
two independent parts, one with the exterior and the other the consequent internal
generation as given by Eq. (\ref{eq: dS eq d{_e}S+d{_i}S}), is difficult to
justify as these constitute a system of interdependent evolutionary interlinked
processes, depending on each other for their sustenance and contribution to
the whole. Thus, {}``life'' forms in which $d\widehat{S}$, arising from the
energy exchanged as food and other sustaining modes with the exterior, is completely
dependent on the capacity $d\widetilde{S}$ of the life to utilise this exchange,
which in turn depends on, and is regulated by $d\widehat{S}$. These interdependent,
non-reductionist, contributions of constituent parts to the whole is a direct
consequence of nonlinearity that effectively implies $f(\alpha x_{1}+\beta x_{2})\neq\alpha f(x_{1})+\beta f(x_{2})$
for the related processes. The second {}``non-life'' example requires the
change to be determined by such internal parameters as mass, specific heat and
chemical concentration of the constituents parts. Thus, for example, in the
adiabatic mixing of a hot and cold body $A$ and $B$ the equilibrium temperature,
given in terms of the respective mole numbers $N$, specific heat $c$ and temperature
$T$, by \begin{equation}
N_{A}c_{A}(T_{A}-T)=N_{B}c_{B}(T-T_{B})\label{eq: equil_temp}\end{equation}
 sets up a state of dynamical equilibrium in which the bi-directional evolutionary
arrow prevents $A$ from annihilating $B$ with the equilibrium condition $T=T_{A}$,
$P=P_{A}$, $\mu=\mu_{A}$. Putting the heat balance equation in the form \[
dQ_{A}+(-dQ_{B})=0,\quad dQ=NcdT\]
suggests that the heat transfer out of a body, considered as a negative real
number, be treated as the additive inverse to the positive transfers into the
system. This sets up a one-to-one correspondence between the forward and its
associated reverse directional naturally occurring real process that evolves
to a state of dynamic equilibrium. 

The noteworthy feature of this evolutionary thermodynamics --- based entirely
on (linear) differential calculus --- is that it reduces the dynamics, as in
Eqs. (\ref{eq: dU_differential}) and (\ref{eq: Gibbs_Entropy}), to a separation
of its governing macroscopic extensive variables, and it is relevant to investigate
the extent to which this decoupling of the motive forces responsible for the
evolution is indeed justifiable for the strongly nonlinear, self-organizing
and emerging complex dynamical systems of nature%
\footnote{\label{foot: baranger}The following extracts from the remarkably explicit lecture
MIT-CTP-3112 by Michel \citet{Baranger2000}, delivered possibly in 2000/2001,
are worth recalling . {}``Chaos is still not part of the American university's
physics curriculum; most students get physics degrees without ever hearing about
it. The most popular textbook in classical mechanics does not include chaos.
Why is that? The answer is simple. Physicists did not have the time to learn
chaos, because they were fascinated by something else. That something else was
20th century physics of relativity, quantum mechanics, and their myriad of consequences.
Chaos was not only unfamiliar to them; it was slightly distasteful!'' 

In offering an explanation for this, Baranger argues that in discovering Calculus,
Newton and Leibnitz {}``provided the scientific world with the most powerful
new tool since the discovery of numbers themselves. The idea of calculus is
simplicity itself. Smoothness (of functions) is the key to the whole thing.
There are functions that are not smooth $\cdots$''. The discovery of calculus
led to that of Analysis and {}``after many decades of unbroken success with
analysis, theorists became imbued with the notion that all problems would eventually
yield to it, given enough effort and enough computing power. If you go to the
bottom of this belief you find the following. \emph{Everything} can be reduced
to little pieces, therefore everything can be known and understood, if we analyze
it to a fine enough scale. The enormous success of calculus is in large part
responsible for the decidedly reductionist attitude of most twentieth century
science, the belief in absolute control arising from detailed knowledge.'' 

Nontheless, {}``chaos is the anti-calculus revolution, it is the rediscovery
that calculus does not have infinite power. Chaos is the collection of those
mathematical truths that have nothing to do with calculus. Chaos theory solves
a wide variety of scientific and engineering problems which do not respond to
calculus.''%
}. Such a separation of variables tacitly implies, as in the example considered
above, that the total energy exchange taking place when the gases are allowed
to mix completely is separable into independent parts arising from changes in
temperature and volume, or from diffusion and mixing of the gases, \emph{with
none of them having any effect on the others.} Recalling that the basic property
of a complex system that serves to define its {}``complex'' character is the
interdependence of its interacting parts responsible for non-reductionism, this
contrary implication of independence of the extensive parameters directly conflicts
with the notions of chaos and complexity. 

The objective of this paper is to propose an explicitly nonlinear, topological
formulation of dynamical evolution in an integrated \emph{chanoxity} --- chaos,
nonlinearity, complexity --- form that focuses on nonlinearity generating self-organization,
adaption, and emergence.

\subsection{\label{sub: Maximal Noninjectivitivity}Maximal Noninjectivity is Chaos}

Chaos was defined in \citet{Sengupta2003} as representing maximal non-injective
ill-posedness in the temporal evolution of a dynamical system and was based
on the purely set theoretic arguments of Zorn's Lemma and Hausdorff Maximal
Chain Theorem. It was, however, necessary to link this with topologies because
evolutionary directions are naturally represented by adherence and convergence
of the associated nets and filters, and this require topologies for describing
their eventual and frequenting behaviour. For this we found the topology of
saturated sets generated by the increasingly non-injective evolving maps (leading
thereby to maximality of ill-posedness and hence to chaos) to provide the motivation
for maximal non-injectivity that in turn leads to the concept of the ininality
of topologies generated by a function $f:\!:X\rightarrow Y$ which is simultaneously
image and preimage continuous. In this case, the topologies on the range $\mathscr{R}(f)$
and domain $\mathscr{D}(f)$ of $f$ are locked with respect to each other as
far as further temporal evolution of $f$ is concerned by having the respective
topologies defined as the $f$-images in $Y$ of $f^{-}$-saturated open sets
of $X$. Thus equation (\ref{eq: bijective_homeo}), taken with $U=\mathscr{S}_{e}(U)$
and $\mathscr{C}_{q}(V)=V$ that are simple consequences of the definitions
\begin{equation}
\textrm{IT}\{ e;\mathcal{V}\}\overset{\textrm{def}}=\{ U\subseteq X\!:U=e^{-}(V),\, V\in\mathcal{V}\}\label{eq: IT}\end{equation}

\noindent and \textsl{}\begin{equation}
\textrm{FT}\{\mathcal{U};q\}\overset{\textrm{def}}=\{ V\subseteq Y\!:q^{-}(V)=U,\, U\in\mathcal{U}\}\label{eq: FT}\end{equation}

\noindent of initial and final topologies, defines $\textrm{for some }V\in\mathcal{V}$
a subset $U\in\mathcal{{U}}$ satisfying \sublabon {equation}\begin{equation}
e^{-}ee^{-}(V)=e^{-}(V):\overset{\textrm{IT}}=U=e^{-}e(U)\Longleftrightarrow(U=\mathscr{S}_{e}(U))\,{\textstyle \bigwedge}\,(e(U)=\mathscr{C}_{e}(V)),\label{eq: IT'}\end{equation}

\noindent while for a subset $A\subseteq U\in\mathcal{U}$ such that $\mathscr{S}_{q}(A)=U$,
the topology $\mathcal{{V}}$ of $Y$ as\begin{equation}
q^{-}q(A)=U:\overset{\textrm{FT}}=q^{-}(V)=q^{-}(qq^{-}(V))\Longleftrightarrow(V=\mathscr{C}_{q}(V))\,{\textstyle \bigwedge}\,(q(U)=V)\label{eq: FT'}\end{equation}
\sublaboff {equation}

\noindent because $qq^{-}=\mathbf{1}_{Y}$ on $\mathscr{R}(q)$; see also column
2, row 1 of Table \ref{tab: ill-posed}. As these equations show, preimage and
image continuous functions are not necessarily open functions: a preimage continuous
function is open iff $e(U)$ is an open set in $Y$ and an image continuous
function is open iff the saturation of every open set of $X$ is also an open
set. The generation of new topologies on the domain and range of a function
--- these will generally be quite different from the original topologies the
spaces might have possessed --- by the evolving dynamics of increasingly nonlinear
maps is a basic property of the evolutionary process that constitutes the motive
for such dynamical changes. Putting these equations together, we get \sublabon {equation}\begin{equation}
U,\, V\in\textrm{IFT}\{\mathcal{U};f;\mathcal{V}\}\Longleftrightarrow(\mathcal{U}=\{ f^{-}(V)\}_{V\in\mathcal{V}})\,{\textstyle \bigwedge}\,(\{ f(U)\}_{U\in\mathcal{U}}=\mathcal{V}),\label{eq: INI}\end{equation}

\noindent which reduces to \begin{equation}
U,\, V\in\textrm{HOM}\{\mathcal{U};f;\mathcal{V}\}\Longleftrightarrow(\mathcal{U}=\{ f^{-1}(V)\}_{V\in\mathcal{V}})\,{\textstyle \bigwedge}\,(\{ f(U)\}_{U\in\mathcal{U}}=\mathcal{V})\label{eq: HOM}\end{equation}
\sublaboff {equation} for a open-continuous bijection $f$ satisfying both
$\mathscr{S}_{f}(A)=A,\,\forall A\subseteq X$ and $\mathscr{C}_{f}(B)=B,\,\forall B\subseteq Y$.
Observe that the only difference between Eqs. (\ref{eq: INI}) and (\ref{eq: HOM})
lies in the one-one and onto character of $f$. 

There are two defining components, temporal and spatial, in any natural evolutionary
processes. However, these are mathematically equivalent in the sense that both
can be represented as pre-ordered sets with the additional directional property
of a \emph{directed set} $(\mathbb{{D}},\preceq)$ which satisfies 

\smallskip{}
(DS1) $\alpha\in\mathbb{{D}\Rightarrow\alpha\preceq\alpha}$ (that is $\preceq$
is reflexive) 

(DS2) $\alpha,\beta,\gamma\in\mathbb{{D}}$ such that $(\alpha\preceq\beta\wedge\beta\preceq\gamma)$
implies $\alpha\preceq\gamma$ (that is $\preceq$ is transitive) 

(DS3) For all $\alpha,\beta\in\mathbb{{D}}$, there exists a $\gamma\in\mathbb{{D}}$
such that $\alpha\preceq\gamma$ and $\beta\preceq\gamma$ 
\smallskip{}

\noindent with respect to the \emph{direction} $\preceq$. While the first two
properties are obvious and constitutes the preordering of $\mathbb{{D}}$, the
third replaces antisymmetry of an order with the condition that every pair of
elements of $\mathbb{{D}}$ always has a successor. This directional property
of $\mathbb{{D}}$, that imparts to the static pre-order a sequential arrow
by allowing it to choose a path of progress between various alternatives that
exists when non-comparable elements bifurcate the arrow, will be used to model
evolutionary processes in space and time. Besides the obvious examples $\mathbb{{N}}$,
$\mathbb{{R}}$, $\mathbb{{Q}}$, or $\mathbb{{Z}}$ of totally ordered sets,
more exotic instances of directed sets imparting directions to neighbourhood
systems in $X$ tailored to the specific needs of convergence theory are summarized
in Table \ref{tab: directions}, where $\beta\in\mathbb{D}$ is the directional
index. %
\begin{table}[htbp]
\noindent \begin{center}{\renewcommand{\arraystretch}{1.5}\begin{tabular}{|c|c|}
\hline 
Directed set $\mathbb{{D}}$&
Direction $\preceq$ induced by $\mathbb{{D}}$\tabularnewline
\hline
\hline 
$_{\mathbb{D}}N=\{ N\!:N\in\mathcal{N}_{x}\}$&
$M\preceq N\Leftrightarrow N\subseteq M$\tabularnewline
\hline 
$_{\mathbb{D}}N_{t}=\{(N,t)\!:(N\in\mathcal{N}_{x})(t\in N)\}$&
$(M,s)\preceq(N,t)\Leftrightarrow N\subseteq M$\tabularnewline
\hline 
$_{\mathbb{D}}N_{\beta}=\{(N,\beta)\!:(N\in\mathcal{N}_{x})(x_{\beta}\in N)\}$&
$(M,\alpha)\leq(N,\beta)\Leftrightarrow(\alpha\preceq\beta)\wedge(N\subseteq M)$\tabularnewline
\hline
\end{tabular}}\end{center}

\caption{{\small \label{tab: directions}Natural directions in $(X,\mathcal{{U}})$
induced by some useful directed sets of convergence theory. Significant examples
of directed sets that are only partially ordered are $(\mathcal{{P}}(X),\subseteq)$,
$(\mathcal{{P}}(X),\supseteq)$; $(\mathcal{{F}}(X),\subseteq)$,$(\mathcal{{F}}(X),\supseteq)$;
$(\mathcal{{N}}_{x},\subseteq)$, $(\mathcal{{N}}_{x},\supseteq)$ for a set
$X$, We take $\mathcal{N}_{x}$, suitably redefined if necessary, to be always
a system of nested subsets of $X$.}}
\end{table}

\noindent While the neighbourhood system $_{\mathbb{D}}N$ at a point $x\in X$
with the \emph{reverse inclusion} direction $\preceq$ is the basic example
of natural direction of the neighbourhood system \emph{}$\mathcal{{N}}_{x}$
of $x$, the more relevant directed sets $_{\mathbb{D}}N_{t}$ and $_{\mathbb{D}}N_{\beta}$
are more convenient in convergence theory because unlike the first, these do
not require a simultaneous application of the Axiom of Choice to every $N\in\mathcal{N}_{x}$.{\large }%
\begin{figure}[htbp]
\noindent \begin{center}{\large \input{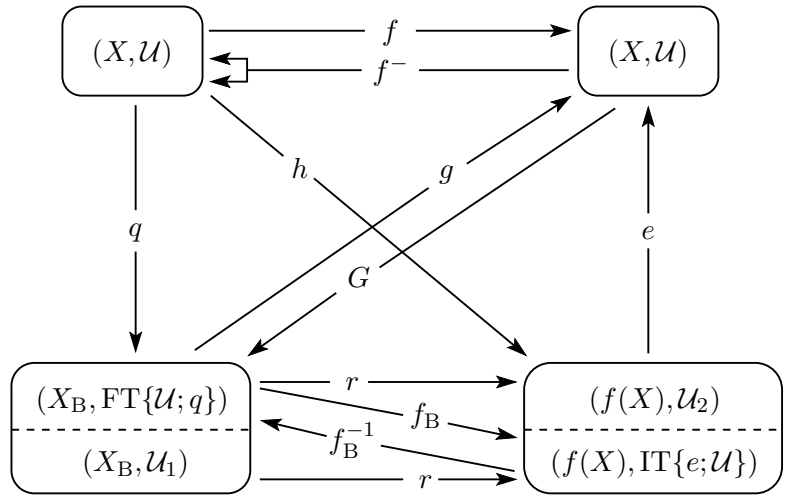}}\end{center}{\large \par}

\caption{\noindent {\large \label{fig: GenInv1}}{\small Generation of a multifunctional
inverse $x=f^{-}(y)$ of the functional equation $f(x)=y$ for $f\!:X\rightarrow X$;
here $G\!:Y\rightarrow X_{\textrm{B}}$ is a generalized inverse of $f$ because
$fGf=f$ and $GfG=G$ that follows from the commutativity of the diagrams. $g$
and $h$ are the injective and surjective restrictions of $f$; these will be
topologically denoted by their generic notations $e$ and $q$ respectively. }}
\end{figure}

Chaos as manifest in the limiting adhering attractors is a direct consequence
of the increasing nonlinearity of the map under increasing iterations and with
the right conditions, appears to be the natural outcome of the characteristic
difference between a function $f$ and its multiinverse $f^{-}$. Equivalence
classes of fixed points stable and unstable\emph{,} as generated by the saturation
operator \emph{}$\mathscr{S}_{f}=f^{-}f$, determine the ultimate behaviour
of an evolving dynamical system, and since the eventual (as also frequent) nature
of a filter or net is dictated by topology on the set, \emph{}chaoticity on
a set \emph{}$X$ \emph{}leads to a reformulation of the open sets of $X$ \emph{}to
equivalence classes generated by the evolving map $f$. In the limit of infinite
iterational evolution in time of $f$ resulting in the multifunction $\Phi$,
the generated open sets constitute a basis for a topology on $\mathscr{D}(f)$
and the basis for the topology of $\mathscr{R}(f)$ are the corresponding $\Phi$-images
of these equivalent classes\emph{.} From the preceding discussions it follows
that the motivation behind the forward evolution of a dynamical system leading
to chaos is the drive toward a state of the dynamical system that supports ininality
of the limit multi \emph{}$\Phi$%
\footnote{\label{foot: ininal}For the logistic map $f_{\lambda}(x)=\lambda x(1-x)$ with
chaos setting in at $\lambda=\lambda_{*}=3.5699456$, this drive in ininality
implies an evolution toward values of the spatial parameter $\lambda\geq\lambda_{*}$;
this is taken to be a spatial parameter as it determines the degree of surjectivity
of $f_{\lambda}$. Together with the temporal evolution in increasing noninjectivity
for any $\lambda$, this comprises the full evolutionary dynamics of the logistic
map. These two distinct dynamical mechanisms of increasing ontoness and increasing
noninjectivity are not independent, however. Thus $\lambda$ --- which we identify
later as representing energy exchanges of all possible types that the system
can have with the surroundings --- determines the nature of the internal forward-backward
stasis that leads to the eventual equilibrium of the system with its environment. %
}. In the limit of infinite iterations therefore, the open sets of the range
$\mathscr{R}(f)\subseteq X$ are the multi images that graphical convergence
generates at each of these inverse-stable fixed points. As readily verified
from Fig. \ref{fig: GenInv1}, $X$ has two topologies imposed on it by the
dynamics of $f$: the first of equivalence classes generated by the limit multi
$\Phi$ in the domain of $f$ and the second as $\Phi$-images of these classes
in the range of $f$. Hence while subdiagrams $X-(X_{\textrm{B}},\textrm{FT}\{\mathcal{{U}};q\})-(f(X),\mathcal{{U}}{}_{2})$
and $(X_{\textrm{B}},\mathcal{{U}}_{1})-(f(X),\textrm{IT}\{ e;\mathcal{{U}}\})-X$
apply to the final and initial topologies of $X_{\textrm{B}}$ and $f(X)$ respectively,
their superposition $X-(X_{\textrm{B}},\textrm{FT}\{\mathcal{{U}};q\})-(f(X),\textrm{IT}\{ e;\mathcal{{U}}\})-X$
\emph{under the additional requirement of a} \emph{homeomorphic} $f_{\textrm{B}}$
leads to the conditions $\mathcal{{U}}_{1}=\textrm{IT}\{ g;\mathcal{{U}}\}$
and $\mathcal{{U}}_{2}=\textrm{FT}\{\mathcal{{U}};h\}$ that $X_{\textrm{B}}$
and $f(X)$ must possess. For this to be possible, {\large }\begin{align*}
\textrm{FT}\{\mathcal{{U}};q\} & =\textrm{IT}\{ g;\mathcal{{U}}\}\\
\textrm{IT}\{ e;\mathcal{{U}}\} & =\textrm{FT}\{\mathcal{{U}};h\}\end{align*}
 requires the image continuous $q$ and the preimage continuous $e$ to be also
be open maps which translates to the ininality of $f$ on $(X,\mathcal{{U}})$,
and hence for the topology of $X$ to be simultaneously the direct and inverse
images of itself under $f$. Recalling that the map $f$ and the topology $\mathcal{{U}}$
of $X$ are already provided, this is interpreted to mean that the increasing
nonlinear ill-posedness of the time-iterates of $f$ is driven by ininality
of the maximally ill-posed limit relation $\Phi$ on $X^{2}$. In this case
$\Phi$ acts as a non-bijective open and continuous relation such that the sequence
of evolving functional relations $(f^{n})$ on $X$ eventually behaves, by Eq.
(\ref{eq: bijective_homeo}), homeomorphically on the saturated open sets of
equivalence classes and their $f^{n}$-images in $X$. We define the resulting
ininal topology on $X$ to be the \emph{chaotic topology} \emph{on} $X$ \emph{associated
with} $f$. Neighbourhoods of points in this topology cannot be arbitrarily
small as they consist of all members of the equivalence class to which any element
belongs; hence a sequence converging to any of these elements necessarily converges
to all of them, and the eventual objective of chaotic dynamics is to generate
a topology in $X$ (irrespective of the original $\mathcal{{U}}$) with respect
to which elements of the space can be grouped together in large equivalence
classes in the sense that if a net converges simultaneously to points $x\neq y\in X$
then $x\sim y$: $x$ is of course equivalent to itself while $x,y,z$ are equivalent
to each other iff they are simultaneously in every open set where the net may
eventually be in. This hall-mark of chaos leads to a necessary eradication of
any separation property that the space might have originally possessed. 

The generation of a new topology on $X$ by the dynamics of $f$ on $X$ is
a consequence of the topology of pointwise biconvergence $\mathcal{T}$ defined
on the set of relations $\textrm{Multi}((X,\mathcal{{U}}),(Y,\mathcal{{V}}))$,
\citep{Sengupta2003}. This generalization of the topology of pointwise convergence
defines neighbourhoods of $f$ in $\textrm{Multi}((X,\mathcal{{U}}),(Y,\mathcal{{V}}))$
to consist of those functions in $(\textrm{Multi}((X,\mathcal{{U}}),(Y,\mathcal{{V}})),\mathcal{T})$
whose images at any point $x\in X$ lie not only close enough to $f(x)\in Y$
(this gives the usual pointwise convergence) but additionally whose inverse
images at $y=f(x)$ contain points arbitrarily close to $x$. Thus the graph
of $f$ must not only lie sufficiently close to $f(x)$ at $x$ in $V\in\mathcal{{V}}$,
but must also be such that $f^{-}(y)$ has at least one branch in the open set
$U\in\mathcal{{U}}$ about $x$. This requires all members of a neighbourhood
$\mathcal{{N}}_{f}$ of $f$ to {}``cling to'' $f$ as the number of points
on the graph of $f$ increases with the result that unlike for simple pointwise
convergence, no gaps in the graph of the limit relation is possible not only
on the domain of $f$ but on its range too. 

For any given integer $I\geq1$, the open sets of $(\textrm{Multi}(X,Y),\mathcal{T})$
are \begin{multline}
B((x_{i}),(V_{i});(y_{i}),(U_{i}))=\{ g\in\mathrm{Map}(X,Y)\!:\\
(g(x_{i})\in V_{i})\,{\textstyle \bigwedge}\,(g^{-}(y_{i})\cap U_{i}\neq\emptyset)\textrm{ },i=1,2,\cdots,I\},\label{eq: func_bi}\end{multline}

\noindent where $(x_{i})_{i=1}^{I}\in X$, $(y_{i})_{i=1}^{I}\in Y$, $(U_{i})_{i=1}^{I}\in\mathcal{{U}}$
$(V_{i})_{i=1}^{I}\in\mathcal{{V}}$ are chosen arbitrarily with reference to
$(x_{i},f(x_{i}))$. A local base at $f$, for $(x_{i},y_{i})\in\textrm{Graph}(f)$,
is the set of functions of (\ref{eq: func_bi}) with $y_{i}=f(x_{i})$, and
the collection of all local bases $B_{\alpha}=B((x_{i})_{i=1}^{I_{\alpha}},(V_{i})_{i=1}^{I_{\alpha}};(y_{i})_{i=1}^{I_{\alpha}},(U_{i})_{i=1}^{I_{\alpha}}),$
for every choice of $\alpha\in\mathbb{D}$, is a base $_{\textrm{T}}\mathcal{B}$
of $(\textrm{Multi}(X,Y),\mathcal{T})$; note that in this topology $(\textrm{Map}(X,Y),\mathcal{T})$
is a subspace of $(\textrm{Multi}(X,Y),\mathcal{T})$. The basic technical tools
needed for describing the adhering limit relation in $(\textrm{Multi}(X,Y),\mathcal{T})$
is the algebraic concept of a filter which is a collection of subsets of $X$
satisfying

\smallskip{}
(F1) The empty set $\emptyset$ does not belong to $\mathcal{F}$, 

(F2) The intersection of any two members of a filter is another member of the
filter: $F_{1},F_{2}\in\mathcal{F}\Rightarrow F_{1}\cap F_{2}\in\mathcal{F}$, 

(F3) Every superset of \emph{}a member of a filter belongs to the filter: $(F\in\mathcal{F})\wedge(F\subseteq G)\Rightarrow G\in\mathcal{F}$;
in particular $X\in\mathcal{F}$, 
\smallskip{}

\noindent and is generated by a subfamily $(B_{\alpha})_{\alpha\in\mathbb{D}}=\,_{\textrm{F}}\mathcal{B}\subseteq\mathcal{F}$
of itself, known as the filter-base, characterized by

\smallskip{}
(FB1) There are no empty sets in the collection $_{\textrm{F}}\mathcal{B}$:
$(\forall\alpha\in\mathbb{D})(B_{\alpha}\neq\emptyset)$

(FB2) The intersection of any two members of \emph{}$_{\textrm{F}}\mathcal{B}$
\emph{}contains another member of $_{\textrm{F}}\mathcal{B}$: $B_{\alpha},B_{\beta}\in\,_{\textrm{F}}\mathcal{B}\Rightarrow(\exists B\in\,_{\textrm{F}}\mathcal{B}\!:B\subseteq B_{\alpha}\cap B_{\beta})$.
\smallskip{}

\noindent Hence any family of subsets of $X$ that does not contain the empty
set and is closed under finite intersections is a base for a unique filter on
$X$, and the filter-base \begin{align}
_{\textrm{F}}\mathcal{B} & \overset{\textrm{def}}=\{ B\in\mathcal{F}\!:B\subseteq F\textrm{ for each }F\in\mathcal{F}\}\label{eq: FB}\end{align}

\noindent \textsl{\emph{determines the filter}} \begin{align}
\mathcal{F} & =\{ F\subseteq X\!:B\subseteq F\textrm{ for some }B\textrm{ }\in\,_{\textrm{F}}\mathcal{B}\}\label{eq: filter_base}\end{align}

\noindent \textsl{\emph{as all the supersets of these basic elements}}\textsl{.}
Note that the filter is an algebraic concept without any topological content;
in order to be able to use it for the purely topological needs of convergence,
a comparison of (F1)-(F3) and (FB1)-(FB2) with (N1)-(N3) and (NB1)-(NB2) of
Sec. \ref{sec: 1. Introduction} show that the neighbourhood system $\mathcal{N}_{x}$
at $x$ is the \emph{neighbourhood filter at} $x$ \emph{}and that any local
base at $x$ is a filter-base for $\mathcal{N}_{x}$: generally for any subset
$A$ of $X$, $\{ N\subseteq X\!:A\subseteq\textrm{Int}(N)\}$ is a filter on
$X$ at $A$. All subsets of $X$ containing a point $p\in X$ is the \emph{principal
filter} $_{\textrm{F}}\mathcal{P}(p)$ \emph{}on $X$ at \emph{}$p$\emph{.}
More generally, the collection of all supersets of a nonempty subset $A$ of
$X$ is the principal filter \emph{}$_{\textrm{F}}\mathcal{P}(A)=\{ N\subseteq X\!:A\subseteq\textrm{Int}(N)\}$
\emph{}at \emph{}$A$\emph{.} The singleton sets $\{\{ x\}\}$ and $\{ A\}$
are particularly simple examples of filter-bases that generate the principal
filters at $\{ x\}$ and $A$; other useful examples that we require subsequently
are the set of all residuals \[
\textrm{Res}(\mathbb{{D}})=\{\mathbb{R}_{\alpha}\!:\mathbb{R}_{\alpha}=\{\beta\in\mathbb{D}\!:\beta\succeq\alpha\in\mathbb{D}\}\}\]
 of a directed set $\mathbb{{D}}$, and the neighbourhood systems $\mathcal{{B}}_{x}$
and $\mathcal{{N}}_{x}$. By \emph{}adjoining the empty set to this filter gives
the $p$-inclusion and $A$-inclusion topologies on $X$ respectively\emph{.} 

The utility of filters in describing convergence in topological spaces arises
from fact that \textsl{}a filter \textsl{$\mathcal{F}$} on \textsl{$X$} can
always be associated with the net \textsl{$\chi_{\mathcal{F}}\!:\,_{\mathbb{D}}F_{x}\rightarrow X$}
defined by \textsl{}\begin{equation}
\chi_{\mathcal{F}}(F,x)\overset{\textrm{def}}=x\label{eq: filter->net}\end{equation}
 \textsl{}where \textsl{$_{\mathbb{D}}F_{x}=\{(F,x)\!:(F\in\mathcal{F})(x\in F)\}$}
is a directed set with direction \textsl{$(F,x)\preceq(G,y)\Rightarrow(G\subseteq F)$;}
reciprocally a net \textsl{$\chi\!:\mathbb{D}\rightarrow X$} corresponds to
the filter-base \begin{equation}
_{\textrm{F}}\mathcal{B}_{\chi}\overset{\textrm{def}}=\{\chi(\mathbb{R}_{\alpha})\!:\textrm{Res}(\mathbb{D})\rightarrow X\textrm{ for all }\alpha\in\mathbb{D}\},\label{eq: net->filter}\end{equation}
 \textsl{}with the corresponding filter \textsl{$\mathcal{F}_{\chi}$} being
obtained by taking all supersets of the elements of \textsl{}$_{\textrm{F}}\mathcal{B}_{\chi}$\textsl{.}
Filters and their bases are extremely powerful tools for maximal non-injective
ill-posedness in the context of the algebraic Hausdorff Maximal Principle and
Zorn's Lemma%
\footnote{\label{foot: Zorn}\textbf{Hausdorff Maximal Principle} (HMP)\textbf{:} Every
partially ordered set has a maximal chain.

\textbf{Zorn's Lemma:} Every inductive set has at least one maximal element. 

A partially ordered set $X$ is said to be \emph{inductive} if every chain of
$X$ has an upper \emph{}bound in $X$. %
}, that we summarize below. 

Let $f$ be a noninjective map in $\textrm{Multi}(X)$ and $\mathscr{I}(f)$
be the number of injective branches of $f$; let \[
F=\{ f\in\textrm{Multi}(X)\!:f\textrm{ is a noninjective function on }X\}\in\mathcal{{P}}(\textrm{Multi}(X))\]
 be the collection of all noninjective functions such that 

\smallskip{}
(1) For every $\alpha$ in a directed set $\mathbb{D}$, $F$ has the extension
property \[
(\forall f_{\alpha}\in F)(\exists f_{\beta}\in F)\!:\mathscr{I}(f_{\alpha})\leq\mathscr{I}(f_{\beta}).\]

\noindent Define a partial order $\preceq$ on $\textrm{Multi}(X)$ for $f_{\alpha},f_{\beta}\in\textrm{Map}(X)\subseteq\textrm{Multi}(X)$
by \begin{equation}
\mathscr{I}(f_{\alpha})\leq\mathscr{I}(f_{\beta})\Longleftrightarrow f_{\alpha}\preceq f_{\beta},\label{eq: Order-Chaos}\end{equation}

\noindent with $\mathscr{I}(f):=1$ for the smallest $f$ denote a partial ordering
$(\textrm{Multi}(X),\preceq)$ of $\textrm{Multi}(X)$. This is actually a preorder
on $\textrm{Multi}(X)$ in which all function with the same number of injective
branches are equivalent to each other. Observe that $\textrm{Multi}(X)$ has
two orders imposed on it: the first $\preceq$ between its elements $f$, and
the second the usual $\subseteq$ that orders subsets of these functional elements.

(2) Let \begin{equation}
C_{\nu}=\{ f_{\alpha}\in\textrm{Multi}(X)\!:f_{\alpha}\preceq f_{\nu}\}\in\mathcal{{P}}(\textrm{Multi}(X)),\qquad\nu\in\mathbb{D},\label{eq: C_nu}\end{equation}
be chains of non-injective functions where $f_{\alpha}\in F$ is to be identified
with the iterates $f^{i}$, the number of injective branches $\mathscr{I}(f)$
depending on $i$. The chains are built from the smallest $C_{0}$, {\small }the
domain $\mathscr{D}$ of $f${\small ,} by application of a choice function
$g_{\textrm{C}}$ that generates the immediate successor \[
C_{j}:=g(C_{i})=C_{i}\,{\textstyle {\textstyle \bigcup}}\, g_{\textrm{C}}(\mathscr{G}(C_{i})-C_{i})\in\mathcal{{X}}\]
 of $C_{i}$ by picking one from the many \[
\mathscr{G}(C_{i})=\{ f\in F-C_{i}\!:\{ f\}\,{\textstyle \bigcup}\, C_{i}\in\mathcal{{X}}\}\]
 that $C_{i}$ may possibly possess; here \begin{equation}
\mathcal{X}=\{ C\in\mathcal{P}(F)\!:C\textrm{ is a chain in }(\textrm{Multi}(X),\preceq)\}\in\mathcal{{P}}^{2}(\textrm{Multi}(X))\label{eq: chi_C}\end{equation}

\noindent is the collection of all chains in $\textrm{Multi}(X)$ with respect
to the order (\ref{eq: Order-Chaos}). Applying $g$ to $C_{0}$ $n$-times
produces the chain $C_{n}=\{\mathscr{D},f(\mathscr{D}),\cdots,f^{n}(\mathscr{D})\}$,
and the smallest common chain\begin{align}
\mathcal{C} & =\{ C_{j}\in\mathcal{P}(\textrm{Multi}(X))\!:C_{i}\subseteq C_{k}\textrm{ for }i\leq k\}\subseteq\mathcal{{X}}\label{eq: chain_chi}\\
 & =\{\mathscr{D},\{\mathscr{D},f(\mathscr{D})\},\{\mathscr{D},f(\mathscr{D}),f^{2}(\mathscr{D})\},\cdots\}\qquad C_{0}:=\mathscr{D}\nonumber \end{align}
 of all the possible $g$-towered chains $\{ C_{i}\}_{i=0,1,2,\cdots}$ of $\textrm{Multi}(X)$
constitutes a principal filter of totally ordered subsets of $(\textrm{Multi}(X),\subseteq)$
at $C_{0}$. Notice that while $\mathcal{{X}}\in\mathcal{{P}}^{2}(\textrm{Multi}(X))$
is a set of sets, $C\in\mathcal{{P}}(\textrm{Multi}(X))$ is relatively simpler
as a set of elements of $f\in\textrm{Multi}(X)$, which at the base level of
the tree of interdependent structures of $\textrm{Multi}(X)$, is canonically
the simplest. {\large }%
\begin{figure}[htbp]
\noindent \begin{center}{\large \input{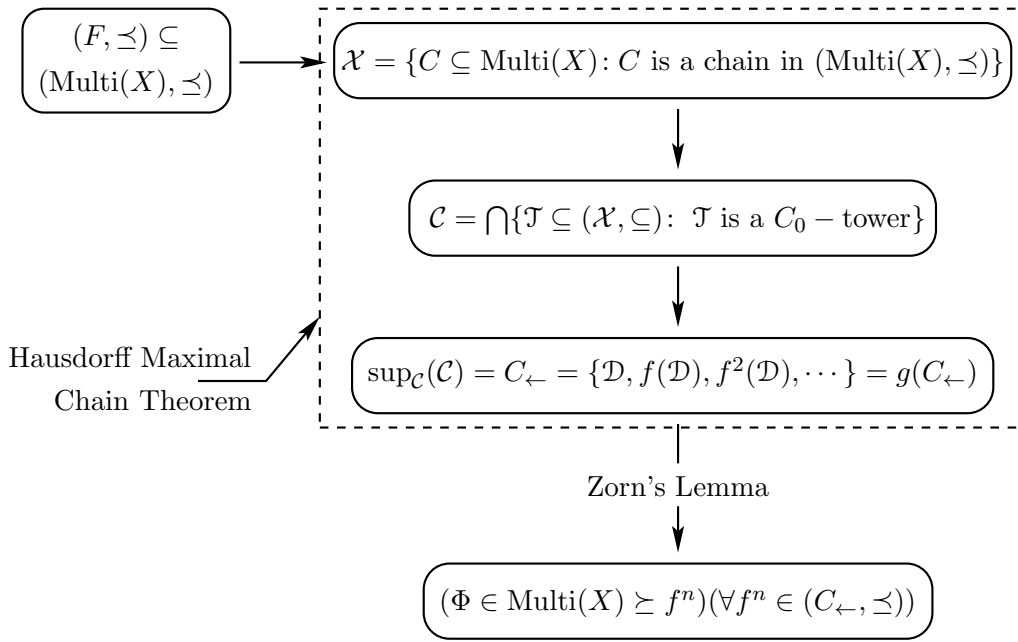}}\end{center}{\large \par}

\caption{{\large \label{fig: Zorn1}}{\small Application of Zorn's Lemma to a partially
ordered set $F=\{ f\in\textrm{Multi}(X)\!:f\textrm{ is a noninjective function}\}$.
$\mathcal{C}=\{\mathscr{D},\{\mathscr{D},f(\mathscr{D})\},\{\mathscr{D},f(\mathscr{D}),f^{2}(\mathscr{D})\},\cdots\}$
is a chain of towered chains of} \emph{\small functions} {\small in $\textrm{Multi}(X)$
with $C_{0}=\mathscr{D}$, the domain of $f$. Notice that to obtain a maximal
$\Phi$ at the base level $\textrm{Multi}(X)$, it is necessary to go two levels
higher: $\mathcal{X}\in\mathcal{{P}}^{2}(\textrm{Multi}(X))\rightarrow C\in\mathcal{{P}}(\textrm{Multi}(X))\rightarrow\Phi\in\textrm{Multi}(X)$
is a three-tiered structure with the two-tiered HMP feeding to the third of
Zorn's Lemma.}}
\end{figure}

To continue further with the application of Hausdorff Maximal Principle to the
partially ordered set $(\mathcal{{X}},\preceq)$ of sets, it is necessary that 

(a) There exists a smallest element $C_{0}$ in $\mathcal{{X}}$ with no predecessor, 

(b) Every element $C\in\mathcal{{X}}$ has an immediate successor $g(C)$ in
$\mathcal{{X}}$ such that there is no element of $\mathcal{{X}}$ lying strictly
between $C$ and $g(C)$, and 

(c) $\mathcal{{X}}$ is an \emph{inductive} set in the sense that every chain
$\mathcal{{C}}$ of $(\mathcal{{X}},\preceq)$ has a supremum $\sup_{\mathcal{{X}}}(\mathcal{{C}})=\cup_{C\in\mathcal{{C}}}C$
in $\mathcal{{X}}$, see footnote \ref{foot: Zorn}. 

Any subset $\mathscr{T}$ of $\mathcal{{X}}$ satisfying these conditions is
often graphically referred to as a \emph{tower}; $\mathcal{{X}}$ is of course
a tower by definition. The intersection of all possible towers of $\mathcal{{X}}$
is the towered chain $\mathcal{{C}}$ of $\mathcal{{X}}$, Eq. (\ref{eq: chain_chi}).
Criterion (c) above is especially crucial as it effectively disqualifies $(F,\preceq)$
as a likely candidate for HMP: the supremum of the chains of increasingly non-injective
functions need not be \emph{}a \emph{function,} but is likely to be a \emph{}multifunction.
Hence $\mathcal{{X}}$ in the conditions above is the space of relations, and
it is necessary to consider $C$ of Eq. (\ref{eq: C_nu}) as a subset of this
$\textrm{Multi}(X)$ rather than of $F$. The careful reader cannot fail to
note that the induction of $\textrm{Multi}(X)$ effectively leads to an {}``extension''
of $\textrm{Map}(X)$ to the set of arbitrary relations wherein the supremum
of the chain of non-injective functions may possibly lie. However it must be
realized that in this purely algebraic setting without topologies on the sets,
the supremum constitutes only a static cap on the family of equilibrium ordered
states: the chains being only ordered and not directed are devoid of any dynamical
evolutionary character. 

(3) Application of the Hausdorff Maximal Principle to $(\mathcal{{X}},\subseteq)$
now yields \begin{align}
\sup_{\mathcal{C}}(\mathcal{C})=C_{\leftarrow} & =\{ f_{\alpha},f_{\beta},f_{\gamma},\cdots\}\nonumber \\
 & =\{\mathscr{D},f(\mathscr{D}),f^{2}(\mathscr{D}),\cdots\}=g(C_{\leftarrow})\in\mathcal{C}\label{eq: sup{_C}(C)}\end{align}
 as the supremum of $\mathcal{C}$ in $\mathcal{C}$, defined as a fixed-point
of the tower generator $g$, without any immediate successor. Identification
of this fixed-point supremum as one of the many possible maximal elements of
$(\mathcal{{X}},\subseteq)$ completes the application of Hausdorff Principle,
yielding $C_{\leftarrow}$ as the required maximal chain of $(\mathcal{{X}},\subseteq)$. 

The technique of HMP is noteworthy because it presents a graphic step-wise algorithmic
rule leading to an equivalent filter description and the algebraic notion of
a \emph{chained tower.} Not possessing any of the topological directional properties
associated with a net or sequence, the tower comprises an ideal mathematical
vocabulary for an ordered succession of equilibrium states of a quasi-static,
reversible, process. The directional attributes of convergence and adherence
must be externally imposed on towered filters like $\mathcal{C}$ by introducing
the neighbourhood system: a filter $\mathcal{{F}}$ converges to $x\in(X,\mathcal{{U}})$
iff $\mathcal{{N}}_{x}\subseteq\mathcal{{F}}$. 

(4) Returning to the partially ordered set $(\textrm{Multi}(X),\preceq)$, Zorn's
Lemma applied to the maximal chained element $C_{\leftarrow}$ of the inductive
set $\mathcal{{X}}$ finally yields the required maximal element $\Phi\in\textrm{Multi}(X)$
as an upper bound of the maximal chain $(C_{\leftarrow},\preceq)$. Because
this limit need not in general be a function, the supremum does not belong to
the towered chain having it as a fixed point, and may be considered as a contribution
of the inverse functional relations $(f_{\alpha}^{-})$ in the following sense.
From Eq. (\ref{eq: func-multifunc}), the net of increasingly non-injective
functions of Eq. (\ref{eq: Order-Chaos}) implies a corresponding net of decreasingly
multivalued functions ordered inversely by the relation $f_{\alpha}\preceq f_{\beta}\Leftrightarrow f_{\beta}^{-}\preceq f_{\alpha}^{-}$.
Thus the inverse relations which are as much an integral part of graphical convergence
as are the direct relations, have a smallest element belonging to the multifunctional
class. Clearly, this smallest element as the required supremum of the increasingly
non-injective tower of functions defined by Eq. (\ref{eq: Order-Chaos}), serves
to complete the significance of the tower by capping it with a {}``boundary''
element that can be taken to bridge the classes of functional and non-functional
relations on $X$.

Having been assured of the existence of a largest element $\Phi\in\textrm{Multi}(X)$,
we now proceed to construct it topologically. Let $(\chi_{i}:=f^{i}(A))_{i\in\mathbb{N}}$
for a subset $A\subseteq X$ that we may take to be the domain of $f$, correspond
to the ordered sequence (\ref{eq: Order-Chaos}). Using the notation of Eq.
(\ref{eq: net->filter}), let the totality of the sequences $\chi(\mathbb{R}_{i})={\scriptstyle \bigcup}_{j\geq i}f^{j}(A)$
for each $i\in\mathbb{N}$ generate the decreasingly nested filter-base \begin{align}
_{\textrm{F}}\mathcal{B} & \overset{\textrm{def}}=\left\{ {\textstyle \bigcup}_{j\geq i}f^{j}(A)\right\} _{i\in\mathbb{{N}}}\nonumber \\
 & =\left\{ {\textstyle \bigcup}_{j\geq i}f^{j}(x)\right\} _{i\in\mathbb{{N}}}\qquad\forall x\in A,\label{eq: absorbing set}\end{align}
 corresponding to the sequence of functional iterates $(f^{j})_{j\geq i\in\mathbb{N}}$.
The existence of a maximal chain with a maximal element guaranteed by the Hausdorff
Maximal Principle and Zorn's Lemma respectively implies a nonempty core of $_{\textrm{F}}\mathcal{B}$.
We now identify this filterbase with the neighbourhood base at $\Phi$ and thereby
define \begin{align}
\Phi(A) & \overset{\textrm{def}}=\textrm{adh}(\,_{\textrm{F}}\mathcal{B})\label{eq: attractor_adherence}\\
 & ={\displaystyle {\textstyle \bigcap}_{i\geq0}\,\textrm{Cl}(A_{i})},\quad A_{i}=\{ f^{i}(A),f^{i+1}(A),\cdots\}\nonumber \end{align}
 as the attractor of $A$, where the closure is with respect to the topology
of pointwise bi-convergence induced by the neighbourhood filter base $_{\textrm{F}}\mathcal{B}$.
Clearly the attractor as defined here is the graphical limit of the sequence
of functions $(f^{i})_{i\in\mathbb{N}}$ with respect to the directed sets of
Table \ref{tab: directions}. This attractor represents, in the product space
$X\times X$, the converged limit of the bi-directional evolutionary dynamics
occurring in the kitchen \emph{}space \emph{}$X\times\mathfrak{{X}}$ (the \emph{anti-space}
$(\mathfrak{{X}},\mathcal{{U}}_{\ominus}){}$ of $(X,\mathcal{{U}})$ is defined
below) that induces the observable image $\Phi(A)$ in $X$. The antispace is
not directly observable, being composed of \emph{anti-elements} $\mathfrak{{x}}$
that correspond in an unique, one-to-one fashion to the corresponding defining
observables $x\in X$, just as the negative reals --- which are not physically
directly observable either --- are attached in a one-to-one fashion with their
corresponding defining positive counterparts in the manner \begin{equation}
r+(-r)=0,\qquad r\in\mathbb{{R}}_{+}.\label{eq: negative}\end{equation}
 The anti-space is necessary for understanding the bi-directional evolutionary
process responsible for a stasis of dynamic equilibrium of two sub-systems competitively
collaborating with each other. The basic example of an anti-space is that of
the negative reals with a \emph{forward} direction in the sense of the \emph{decreasing}
negatives resulting from an \emph{exclusion} anti-\emph{topology} $\mathcal{{U}}_{\ominus}$
which is generated by the topology $\mathcal{{U}}$ of the observable positive
reals $R_{+}$. This generalization of the additive inverse of the real number
system to sets is considered in the next subsection.

\subsubsection{\label{sub: Antispace}The Antispace of a topological space}

\textbf{Postulate 1. The anti-set $\mathfrak{{X}}$.}%
\footnote{\label{foot: anti}Anti-sets will be denoted by $\mathfrak{{fraktur}}$ letters. %
} Let $X$ be a set and suppose that for every $x\in X$ there exists an $\mathfrak{{x}\in\mathfrak{{X}}}$
with the property that\sublabon{equation}\begin{eqnarray}
\mathfrak{{X}} & \!\!\overset{\textrm{def}}=\!\! & \{\mathfrak{{x}}\!:\{ x\}\,{\textstyle \bigcup}\,\{\mathfrak{{x}}\}=\emptyset\}\label{eq: matter-anti(a)}\end{eqnarray}
 defines the \emph{anti-set} (also to be referred to as the \emph{anti-image})
of $X$. This means that for every subset $A$ of $X$ there is an anti-set
$\mathfrak{{A}}\subseteq\mathfrak{{X}}$ \emph{associated with} (\emph{generated
by}) it such that \begin{eqnarray}
A\,{\textstyle \bigcup}\,\mathfrak{{B}} & \!\!\overset{\textrm{def}}=\!\! & A-B,\qquad B\longleftrightarrow\mathfrak{{B}},\label{eq: matter-anti(b)}\end{eqnarray}
\sublaboff{equation}implies $A\cup\mathfrak{{A}}=\emptyset$. Hence anti-sets
of $X$ act as \emph{inhibitors} or \emph{moderators} of $X$. 
\smallskip{}

As compared with the directed set $(\mathcal{{P}}(X),\subseteq)$ that induces
the natural direction of \emph{decreasing} \emph{subsets} of Table \ref{tab: directions},
the direction of \emph{increasing} \emph{supersets} induced by $(\mathcal{{P}}(X),\supseteq)$
--- which understandably finds no ready application in convergence theory ---
is useful in generating an anti-topology $\mathcal{{U}}_{\ominus}$ on $\mathfrak{{X}}$
corresponding to the topology $\mathcal{{U}}$ in $X$ as follows. Let $(x_{0},x_{1},x_{2},\cdots)$
be a sequence in $X$ converging to $x_{*}\in X$ with reference to any of the
reverse inclusion, forward direction, of decreasing neighbourhood system $\mathcal{{N}}_{x_{*}}$
of Table \ref{tab: directions}, and consider the backward direction induced
at the limit $x_{*}$ by the directed set $(\mathcal{{P}}(X),\supseteq)$ of
increasing supersets containing $x_{*}$. As the reverse sequence $(x_{*},\cdots,x_{i+1},x_{i},x_{i-1},\cdots)$
does not converge to $x_{0}$ unless it is eventually in every neighbourhood
of this initial point, we employ the closed-open subsets {\renewcommand{\arraystretch}{1.15}\begin{align}
N_{i}-N_{j} & =\left\{ \begin{array}{l}
(N_{i}-N_{j})\,{\textstyle \bigcap}\, N_{i},\quad\textrm{(open)}\\
(N_{i}-N_{j})\,{\textstyle \bigcap}\,(X-N_{j})\quad\textrm{(closed)}\end{array}\right.\label{eq: clopen}\end{align}
}$(j>i)$ in the \emph{inclusion} topology of $X$ with $x_{i}\in N_{i}-N_{i+1}$,
$N_{i}\in\mathcal{{N}}_{x_{*}}$, to generate a topology in $\mathfrak{{X}}$
in Postulate 2 below. For this, recall that while the $x$-\emph{inclusion}
\emph{topology} of $X$ comprise all subsets of $X$ that \emph{include} $x$
(together with $\emptyset$) with the neighbourhood system $\mathcal{{N}}_{x}$
being just these non-empty subsets of $X$, the $x$-\emph{exclusion} \emph{topology}
are all those subsets $\mathcal{{P}}(X-\{ x\})$ of $X$ that \emph{exclude}
$x$ (together with $X$), with $\{ X\}$ and $\{\{ y\}\}$ being its neighbourhood
systems at $x$ and any $y\neq x$. Thus while a net trivially converges to
$x$ in its exclusion topology, it converges to any other point $y\neq x$ iff
it is eventually constant at $\{ y,y,y,\cdots\}$%
\footnote{\label{foot: JosephLo}I thank Joseph Lo for his clarifications on the subtleties
of the exclusion topology, Private Communication, May 2004. %
}. Since the open sets of the second of Eq. (\ref{eq: clopen}) in the $x_{*}$-exclusion
topology are actually closed with respect to the inclusion topology, and arbitrary
(respectively finite) union of open (respectively closed) sets belong to their
respective classes, we postulate with respect to the directed set $_{\mathbb{{D}}}N_{i}=\{(N_{i},i)\!:(N_{i}\in\mathcal{{N}}_{x_{*}})(x_{i}\in N_{i})\}$
of Table \ref{tab: directions} and a sequence $(x_{i})_{i\geq0}$ in $(X,\mathcal{{U}})$
converging to $x_{*}=\textrm{adh}_{i\geq0}(\textrm{Cl}(N_{i}))\in X$, that 

\smallskip{}
\noindent \textbf{Postulate 2. The anti-topology $\mathcal{{U}}_{\ominus}$.}
There exists a \emph{decreasing} sequence of moderating anti\emph{-}elements
\emph{}$(\mathfrak{{x}}_{i})_{i<\infty}$ in $\mathfrak{{X}}$ that converges
to $\mathfrak{{x}}_{0}$ in the $\mathfrak{{x}}_{*}$-exclusion topology \textbf{$\mathcal{{U}}_{\ominus}$}
of $\mathfrak{{X}}$ generated by the closed sets $\{ N_{i}-N_{i+1}\}_{i\geq0}$
of $(X,\mathcal{{U}})$, by being eventually constant in the open set $\mathfrak{{N}}_{0}-\mathfrak{{N}}_{1}\in\mathcal{{U}}_{\ominus}$
with value $\mathfrak{{x}}_{0}$; alternatively, all distinct points of these
open sets of $\mathcal{{U}}_{\ominus}$ are equivalent with respect to the converging
sequences. Since the only manifestation of anti-sets in the observable real
world is their inhibitory property, the \emph{decreasing} sequence $(\mathfrak{{x}}_{i})_{i<\infty}$
will be taken to converge to $\mathfrak{{x}}_{0}$ in $(\mathfrak{{X}},\mathcal{{U}}_{\ominus})$
if and only if the \emph{increasing} sequence $(x_{i})_{i\geq0}$ converges
in $(X,\mathcal{{U}})$, that is if and only if the moderating sequence $(\cdots,\mathfrak{{x}}_{(j+1)},\mathfrak{{x}}_{j},\mathfrak{{x}}_{(j-1)},\cdots,\mathfrak{{x}}_{0})$
of $\mathfrak{{X}}$ is eventually in every $\mathcal{{U}}_{\ominus}$-neighbourhood
of $\mathfrak{{x}}_{0}$ as generated by the $\mathcal{{U}}$-closed, $x_{*}$-\emph{exclusion}
open sets, $\{ N_{i}-N_{i+1}\}_{i\geq0}$ of $X$. This decreasing \emph{natural
forward direction} in $(\mathfrak{{X}},\mathcal{{U}}_{\ominus})$ of Table \ref{tab: anti-directions}
is to be compared with the \emph{natural reverse directions} in $(X,\mathcal{{U}})$,
Table \ref{tab: directions}. %
\begin{table}[htbp]
\noindent \begin{center}{\renewcommand{\arraystretch}{1.5}\begin{tabular}{|c|c|}
\hline 
Directed set $\mathbb{{D}}$&
Direction $\preceq$ induced by $\mathbb{{D}}$\tabularnewline
\hline
\hline 
$_{\mathbb{D}}\mathfrak{{N}}=\{\mathfrak{{N}}\!:\mathfrak{{N}}\in\mathcal{N}_{\mathfrak{{x}}}\}$&
$\mathfrak{{M}}\preceq\mathfrak{{N}}\Leftrightarrow\mathfrak{{M}}\subseteq\mathfrak{{N}}$\tabularnewline
\hline 
$_{\mathbb{D}}\mathfrak{{N}}_{\mathfrak{{t}}}=\{(\mathfrak{{N}},\mathfrak{{t}})\!:(\mathfrak{{N}}\in\mathcal{N}_{\mathfrak{{x}}})(\mathfrak{{t}}\in\mathfrak{{N}})\}$&
$(\mathfrak{{M}},\mathfrak{{s}})\preceq(\mathfrak{{N}},\mathfrak{{t}})\Leftrightarrow\mathfrak{{M}}\subseteq\mathfrak{{N}}$\tabularnewline
\hline 
$_{\mathbb{D}}\mathfrak{{N}}_{\beta}=\{(\mathfrak{{N}},\beta)\!:(\mathfrak{{N}}\in\mathcal{N}_{\mathfrak{{x}}})(\mathfrak{{x}}_{\beta}\in\mathfrak{{N}})\}$&
$(\mathfrak{{M}},\alpha)\leq(\mathfrak{{N}},\beta)\Leftrightarrow(\alpha\preceq\beta)\wedge(\mathfrak{{M}}\subseteq\mathfrak{{N}})$\tabularnewline
\hline
\end{tabular}}\end{center}

\caption{{\small \label{tab: anti-directions}Natural forward directions in the antispace
$(\mathfrak{{X}},\mathcal{{U}}_{\ominus})$ is to be compared with Table \ref{tab: directions}
of the natural reverse directions in $(X,\mathcal{{U}})$. The direction of
anti-events in $\mathfrak{{X}}$ is opposite to that of $X$ in the sense that
the temporal sequence of images of events in $X$ opposes that in $\mathfrak{{X}}$
and the order of occurrence of events induced by the anti-world appear to be
reversed to the real observer stationed in $X$. }}
\end{table}

Although the backward sequence $(x_{i})_{i=\cdots,i+1,i,i-1,\cdots}$ in $(X,\mathcal{{U}})$
does not converge, the effect of the containing sequence $(\mathfrak{{x}}_{i})_{i<\infty}$
of $\mathfrak{{X}}$ on $X$ is to inhibit the evolution of the forward sequence
$(x_{i})_{i\geq0}$ to an effective state of dynamical stasis of equilibrium.
It is to be noted that the uni-directional forward arrow $(\mathfrak{{x}}_{*},\cdots,\mathfrak{{x}}_{2},\mathfrak{{x}}_{1},x_{0},x_{1},x_{2},\cdots,x_{*})$
powered by ininality in the composite real-anti world, translates into the bidirectionally
of Eq. (\ref{eq: matter-anti(b)}) responsible for the dynamical stasis. The
significance of these concepts can be appreciated by considering for $X$ and
$\mathfrak{{X}}$ the sets of positive and negative reals, and for $x_{*}$,
$\mathfrak{{x}}_{*}$ a positive real number and its negative inverse image. 

An open set of $X$ is by definition a subset in which a net must eventually
reside in order to converge to a point in that set. The existence of an anti-element
$x\leftrightarrow\mathfrak{{x}}$ in $\mathfrak{{X}}$ for every $x\in X$ requires
all \emph{}forward \emph{increasing} directions in $X$ to have a matching \emph{}forward
\emph{decreasing} direction in $\mathfrak{{X}}$ that actually \emph{appears}
\emph{increasing forward when viewed from} $X$. It is this opposing complimentary
inhibitory effects of $\mathfrak{{X}}$ forward decreasing sequences on $X$
--- responsible by Eq. (\ref{eq: matter-anti(b)}) for moderating the normal
uni-directional evolution in $X$ --- that leads to a stasis of dynamical balance
between the opposing forces generated in the composite of a system with its
environment. Obviously, the evolutionary process will cease when the opposing
influences in $X$ due to itself and that generated by its inhibitor $\mathfrak{{X}}$
balance each other which is the state of dynamic equilibrium. %
\begin{table}[htbp]
\noindent \begin{center}{\renewcommand{\arraystretch}{1.25}\begin{tabular}{|c|c|c|}
\hline 
Property&
$(X,\mathcal{{U}})$&
$(\mathfrak{{X}},\mathcal{{U}}_{\ominus})$\tabularnewline
\hline
\hline 
\multirow{2}{*}{$T_{0}$}&
$(\forall x\neq y\in X)\,(\exists\, N\in\mathcal{{N}}_{x}\!:N\cap\{ y\}$&
$(\forall\mathfrak{{x}}\neq\mathfrak{{y}}\in\mathfrak{{X}})\,(\not\exists\,\mathfrak{{N}}\in\mathcal{{N}}_{\mathfrak{{x}}}\!:\mathfrak{{N}}\cap\{\mathfrak{{y}}\}$\tabularnewline
&
$=\emptyset)\vee(\exists\, M\in\mathcal{{N}}_{y}\!:M\cap\{ x\}=\emptyset)$&
$=\emptyset)\vee(\not\exists\,\mathfrak{{M}}\in\mathcal{{N}}_{\mathfrak{{y}}}\!:\mathfrak{{M}}\cap\{\mathfrak{{x}}\}=\emptyset)$\tabularnewline
\hline 
\multirow{2}{*}{$T_{1}$}&
$(\forall x\neq y\in X)\,(\exists\, N\in\mathcal{{N}}_{x}\!:N\cap\{ y\}$&
$(\forall\mathfrak{{x}}\neq\mathfrak{{y}}\in\mathfrak{{X}})\,(\not\exists\,\mathfrak{{N}}\in\mathcal{{N}}_{\mathfrak{{x}}}\!:\mathfrak{{N}}\cap\{\mathfrak{{y}}\}$\tabularnewline
&
$=\emptyset)\wedge(\exists\, M\in\mathcal{{N}}_{y}\!:M\cap\{ x\}=\emptyset)$&
$=\emptyset)\wedge(\not\exists\,\mathfrak{{M}}\in\mathcal{{N}}_{\mathfrak{{y}}}\!:\mathfrak{{M}}\cap\{\mathfrak{{x}}\}=\emptyset)$\tabularnewline
\hline 
\multirow{2}{*}{$T_{2}$}&
$(\forall x\neq y\in X)\,(\exists\, N\in\mathcal{{N}}_{x}\wedge M\in\mathcal{{N}}_{y})$&
$(\forall\mathfrak{{x}}\neq\mathfrak{{y}}\in\mathfrak{{X}})\,(\not\exists\,\mathfrak{{N}}\in\mathcal{{N}}_{\mathfrak{{x}}}\wedge\mathfrak{{M}}\in\mathcal{{N}}_{\mathfrak{{y}}})$\tabularnewline
&
$:(M\cap N=\emptyset)$&
$:(\mathfrak{{M}}\cap\mathfrak{{N}}=\emptyset)$\tabularnewline
\hline
\end{tabular}}\end{center}

\caption{\label{tab: separation}{\small Comparison of the separation properties of
$(X,\mathcal{{U}})$ and its inhibiting anti-space $(\mathfrak{{X}},\mathcal{{U}}_{\ominus})$.}}
\end{table}

It should be noted that the moderating image $\mathfrak{{X}}$ of $X$ needs
to be endowed with inverse inhibitory properties if (\ref{eq: matter-anti(b)})
is to be meaningful which leads to the separation properties of the conjugate
spaces $(X,\mathcal{{U}})$ and $(\mathfrak{{X}},\mathcal{{U}}_{\ominus})$
shown in Table \ref{tab: separation}. It is significant that the anti-space
is topologically distinguished in having its sequences converge with respect
to increasing neighbourhoods of the limit point, a property that leads as already
pointed out earlier to the existence of a multiplicity of equivalent limits
in large neighbourhoods of $\mathfrak{{x}}_{0}$ to which the sequence in $\mathfrak{{X}}$
converges, even when $(X,\mathcal{{U}})$ is Hausdorff. We conjecture, in the
context of iterational evolution of functions that concerns us here, that the
function-multifunction asymmetry of (\ref{eq: func-multifunc}) introduced by
the non-injectivity of the iterates is directly responsible for the difference
in the separation properties of $\mathcal{{U}}$ and $\mathcal{{U}}_{\ominus}$,
which in turn prohibits the system from annihilating $B$ mentioned earlier
and forces it to adopt the forward-backward stasis of dynamic equilibrium. Recalling
that non-injectivity of one-dimensional maps translate to pairs of injective
branches with \emph{positive} and \emph{negative} slopes, we argue in the context
of Fig. \ref{fig: dynamics} that whereas branches with positive slope represent
matter, those with negative slope correspond to anti-matter by Eq. \ref{eq: matter-anti(b)}.
\begin{figure}[htbp]
\noindent \begin{center}\input{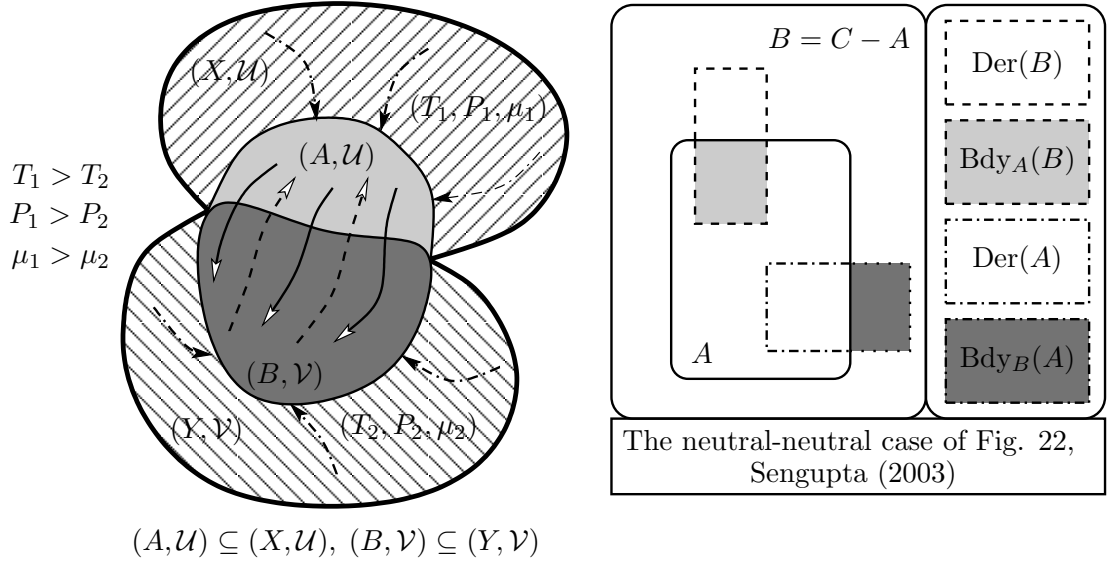}\end{center}

\caption{\label{fig: stasis}{\small Schematic representation of irreversible entropy
generation in $C=A\cup B$ with respect to the universe $X\cup Y$. The irreversible
process is indicated by the nets of full arrows with open heads from $A$ to
$B$} r{\small epresenting transfer of energy, volume, or mass driven by appropriate
evolutionary directed set of a thermodynamic force (for instance due to a temperature
gradient $T_{1}>T_{2}$ inducing energy transfer) that provides the driving
impetus of ininality for the directional transport. The dashed open arrows show
the reverse evolution in $C$ due to its inhibitor $\mathfrak{{C}}$, where
$C\cup\mathfrak{{C}}=\emptyset$. The dash-dot arrows stand for the uni-directional
transfer of energy from a reservoir that continues till the respective parts
of $C$ acquire the characteristics of their reservoirs. We shall identify the
solid arrows in $C$ with second law entropic emergence and the dashed arrows
as anti-entropic self-organization.}}
\end{figure}

As an example of the application of these ideas, let us return to Eqs. (\ref{eq: d{e_}S})
and (\ref{eq: d{i_}S}) for the entropy change due to external exchange and
non-linear, irreversible, internal generation respectively. The external exchange
of energy with the environment leads to a change in the internal state of the
system which is then utilized in performing irreversible useful work relative
to the environment. The situation is conveniently displayed in terms of the
\emph{neutral-neutral} convergence mode of a net schematically represented in
Fig. \ref{fig: stasis} and adapted from Fig. 22 of \Citet*{Sengupta2003},
illustrating the irreversible internal generation of entropy in a universe $C=A\cup B$,
where $A$ and $B$ are two parts of a system prepared at different initial
conditions as shown in the figure. In order to examine these questions in the
evolutionary perspective, we first formalize the notion of 

\smallskip{}
\noindent \textbf{Definition.} \textbf{Interaction} \textbf{between} \textbf{two}
\textbf{spaces}. A space $(A,\mathcal{{U}})$ will be said to \emph{interact}
with a \emph{disjoint} space $(B,\mathcal{{V}})$ if there exists a function
$f$ on the \emph{sum space} $(C,\mathcal{{W}})$, where $C=A\cup B$ and \begin{align*}
\mathcal{{W}} & =\{ W:=U\,{\textstyle \bigcup}\, V\!:(U\in\mathcal{{U}})\,{\textstyle \bigwedge}\,(V\in\mathcal{{V}})\}\\
 & =\{ W\subseteq C\!:(W\,{\textstyle \bigcap}\, A\textrm{ is open in }A)\,{\textstyle \bigwedge}\,(W\,{\textstyle \bigcap}\, B\textrm{ is open in }B)\},\end{align*}
 which evolves graphically to a well defined limit relation in the topology
of pointwise biconvergence on $(C,\mathcal{{W}})$. The function $f$ will be
said to be an \emph{interaction} between $A$ and $B$.%
\footnote{\noindent \label{foot: sum-topology}If $A$ and $B$ are not disjoint, then
this construction of the sum may not work because $A$ and $B$ will generally
induce distinct topologies on $C$; in this case $\mathcal{{W}}$ is obtained
as follows. Endow the disjoint copies $A_{1}:=A\times\{1\}$ and $B_{2}:=B\times\{2\}$
of $A$ and $B$ with topologies $\mathcal{{U}}_{1}=\{ U\times\{1\}\!:U\in\mathcal{{U}}\}$
and $\mathcal{\mathcal{{V}}}_{2}=\{ V\times\{2\}\!:V\in\mathcal{{V}}\}$, which
are homeomorphic with their originals with $a\mapsto(a,1)$ and $b\mapsto(b,2)$
being the respective homeomorphisms. Then $C=A_{1}\cup B_{2}$ is the sum of
$A_{1}$ and $B_{2}$ with the topology $\mathcal{{W}}=\{ W\subseteq C\!:W=(U\times\{1\})\cup(V\times\{2\})\!:(U\in\mathcal{{U}})\wedge(V\in\mathcal{{V}})\}$
inducing the subspaces $(A_{1},\mathcal{{U}}_{1})$ and $(B_{2},\mathcal{\mathcal{{V}}}_{2})$. %
} 
\smallskip{}

The forward evolution in $(C,\mathcal{{W}})$ motivated by the inducement of
an ininal topology on $C$ is opposed by the restraining, inhibiting, and backward
influence arising from the exclusion topologies of the antispace \[
(\mathfrak{{C}},\mathcal{{W}}_{\ominus})=(\mathfrak{{A}}\cup\mathfrak{{B}},\,\mathcal{{U}}_{\ominus}\cup\mathcal{{V}}_{\ominus}),\]
 with the equivalence classes generated in the anti-space being responsible
for the multiinverses of the evolving $f$ that characterises the nonlinear
state of $C$ following the internal preparation of the system. This irreversible
process is indicated in Fig. \ref{fig: stasis} by the nets of open-headed full
arrows from $(A,\mathcal{{U}})$ to $(B,\mathcal{{V}})$ representing transfer
of energy, volume, or mass driven by an appropriate evolutionary directed set
of a thermodynamic force (for instance due to a temperature gradient $T_{A}>T_{B}$
inducing the energy transfer) that provides the driving impetus for directional
transport motivated by ininality. 

Since physical evolution powered by changes in the internal intensive parameters
is represented by convergence of appropriate sequences and nets, it is postulated
in keeping with the role of ininality, that equilibrium in uni-directional temporal
evolutions like $X\rightarrow A\subseteq X$ or $Y\rightarrow B\subseteq Y$
sets up $A$ and $B$ as subspaces of $X$ and $Y$ respectively. For bi-directional
processes like $A\leftrightarrow B$, the open headed dashed arrows of Fig.
\ref{fig: stasis} from $B$ to $A$ represent the inhibiting backward influence
of $(\mathfrak{{C}},\mathcal{{W}}_{\ominus})$ on $(C,\mathcal{{W}})$. The
assumptions 

\smallskip{}
(a) Both the subsets $A$ and $B$ of $C$ are perfect in the sense that $A=\textrm{Der}(A)$
and $B=\textrm{Der}(B)$ so that all points of each of these sets can be reached
by sequences eventually in them, and 

(b) $\textrm{Bdy}_{B}(A)=B$ and $\textrm{Bdy}_{A}(B)=A$ which enables all
points of $A$ and $B$ to be directly accessed as limits by sequences in $B$
and $A$, 
\smallskip{}

\noindent imply that any exchange of energy from the environment $E=X\cup Y$
to system $C$ will be evenly dispersed through it by the irreversible, internal
evolution of the system, once $C$ attains equilibrium with $E$ and is allowed
to evolve unperturbed thereafter. \textbf{}This \emph{global homogenizing} \emph{principle}
\emph{of} \emph{detailed} \emph{balance}, applicable to evolutionary processes
at the micro-level provides a rationale for equilibration in nature that requires
every forward direct process to be balanced by an oppositely directed arrow,
leading to the global \emph{}equilibrium of thermodynamics. If backward influences
exactly balance the inducing forward impetus resulting in a complete restoration
of all the intermediate stages, then the resulting \emph{reversible process}
is actually quasi-static with no effective changes; \emph{}note that nontrivial
equilibrated stasis cannot be generated by any reversible processes. Rather
than subscribing to the additive (linear) decomposition of an environmental
uni-directional energy exchange and the attendant bi-directional internal evolution
implied by Eq. (\ref{eq: dS eq d{_e}S+d{_i}S}), we instead adopt the point
of view that these processes are interrelated and the drive toward ininality
is accompanied by its subsequent internal utilization, gainfully or otherwise
with dissipation. In this sense, interaction will always imply the couple $(f,\mathfrak{{f}})$
of a function $f$ and its anti-self $\mathfrak{{f}}$ rather than $f$ alone. 

With reference to evolution of maps like the logistic $f_{\lambda}=\lambda x(1-x)$,
which for a particular $\lambda$ can be taken to represent the subspace $C\subseteq E$
at equilibrium with its environment $E$, evolutionary changes in $\lambda$
induce changes in the internal intensive thermodynamic parameters that follow
uni-directional exchanges of $C$ with $E$. This perturbs the equilibrium between
components $A$ and $B$ resulting in further evolutionary iterational interaction
between them. The forward iterational evolution of $f_{\lambda}$ is hindered
by the backward restraining effect of $\mathfrak{{C}}$ which suppresses the
continual increase of noninjectivity of $f_{\lambda}$ that would otherwise
lead to a state of maximum noninjective ill-posedness for this $\lambda$. Measurable
global equilibrium represents a balance between the opposing induced local evolutionary
forces that are determined by, and which in turn determine, the degree of energy
exchange $\lambda$. The eventual ininality at $\lambda=4$ represents continual
energy absorption from $E$ that is dissipated for the globalizing uniformity
of Figs. \ref{fig: dynamics} and \ref{fig: Complex-Chaos}(c). For $3<\lambda\leq\lambda_{*}=3.5699456$
the energy input is gainfully employed to generate the complex structures that
are needed to sustain the process at that level of $\lambda$. 

Recalling footnote \ref{foot: ininal}, we now summarize the principal features
of the nonlinear evolutionary dynamics following interaction of a composite
system with its surroundings. 

\smallskip{}
(a) If the state of dynamic equilibrium of a composite system $C=A\cup B$ with
its surroundings, as represented by the logistic map is disturbed by some form
of communication between the two, forces are set up between the components $A$
and $B$ so as to absorb the effect of this disturbance. 

(b) The consumption of the effects of this exchange is motivated by a \emph{simultaneous},
non-reductionist drive towards increasing surjectivity \emph{and} increasing
noninjectivity of the map $f_{\lambda}$ and its evolved iterated images, which
eventually leads to a state of maximal non-injectivity on the domain of $f$.
Owing to the function-multifunction asymmetry of the map, such a condition would
signify static equilibrium and an end to all further evolutionary processes,
a state of dissipative annihilation, burn-out and ininality. 

(c) Since such eventual self-destruction cannot be the stated objective of Nature,
this unrelenting march toward collapse is restrained by the anti-world effects
we have described earlier. Since the anti-world moderates the real, a reversed
sequential direction effectively inhibits the drive towards self-destruction
motivated by the simultaneous increase of $\lambda$ and the increased noninjectivity
of forward iterations, and the resulting state of dynamic equilibrium is the
observed equilibrium of Nature. Like all others, nature's \emph{kitchen} $C\times\mathfrak{{C}}$
where the actual dynamical processes occur is beyond direct observation; only
its moderating effect in \emph{}$C\times C$ is perceived by the observer in
$\mathscr{D}(f)=C$. 
\smallskip{}

As an example of this line of reasoning, consider an isolated system of two
parts with each locally in equilibrium with its environment as in Eq. (\ref{eq: d{i_}S})
that can now be re-expressed as \begin{multline}
\widetilde{S_{C}}(t)=\widetilde{S_{C0}}+\left[N_{A}c_{A}\ln\left(\frac{T}{T_{A}(t)}\right)+N_{B}c_{B}\ln\left(\frac{T}{T_{B}(t)}\right)\right]-\\
-R\left[N_{A}\ln\left(\frac{P_{A}}{p_{A}(t)}\right)+N_{B}\ln\left(\frac{P_{B}}{p_{B}(t)}\right)\right],\label{eq: S_{int}(t)}\end{multline}
where we note with reference to Fig. \ref{fig: stasis} that $T_{A}=T_{1}$,
$T_{B}=T_{2}$ are the temperatures of subsystems $A$ and $B$, $V_{A}+V_{B}=V$
is the total volume of $C$, $p_{A},\, p_{B}$ are the pressures of $A$ and
$B$, $P_{A,B}:=N_{A,B}RT_{A,B}/V$ are their partial pressures with $P=P_{A}+P_{B}$
the total pressure exterted by the gases in $V$, and $T$ is the equilibrium
temperature of (\ref{eq: equil_temp}). 

Then 

(i) If the halves containing nonidentical ideal gases at different temperatures
are brought in contact with each other, the equilibrium state of stasis resulting
from the flow of \emph{heat} and \emph{cold} ($=$ anti-heat) between the bodies
lead to the equality of temperature, $T_{A}=T=T_{B}$, leading to the vanishing
of the first part of Eq. (\ref{eq: S_{int}(t)}). 

(ii) If the gas in the first half expands into the second then equilibrium is
reached when the \emph{gas} \emph{outflow} from the first half into the second
is exactly balanced by the \emph{vacuum inflow} from the second into the first
if the second is evacuated, or if filled with a nonidentical gas then equalization
of pressure of the chambers by outflow of the gases from their respective halves
into the other, results in the vanishing of the second term of (\ref{eq: S_{int}(t)}).
In either of these cases competitive collaboration of the two halves, one with
greater resources than the other, rather than annihilation of the weaker by
the more resourceful leads to the state of mutual equilibrium.

In all these instances, the effect of the antiworld on the real is to moderate,
inhibit or contain the consequence of the latter: this is its only manifestation
in the observable real world. Thus \emph{cold, vacuum} and \emph{a nonidentical
substance} are the negations of \emph{heat} and \emph{matter ---} just as $-r\in\mathbb{{R}}_{-}$
is the negation of $r\in\mathbb{{R}}_{+}$\emph{.} These negations as elements
of the anti-world are no more observable than $-5$, for example, is to us in
our real world: we cannot have $-5$ objects around us, or measure the distance
between two places to be $-100$ kilometers. Nonetheless, without $\mathbb{{R}}_{-}$
there would be no zero, no starting initial point in any ordered set, and no
{}``equilibrium'' either. Nature, propelled by the unidirectional increase
in entropic disorder, without the moderating influence induced by its anti-self,
would have possibly crashed out of existence long ago! %
\begin{figure}[htbp]
\noindent \begin{center}\input{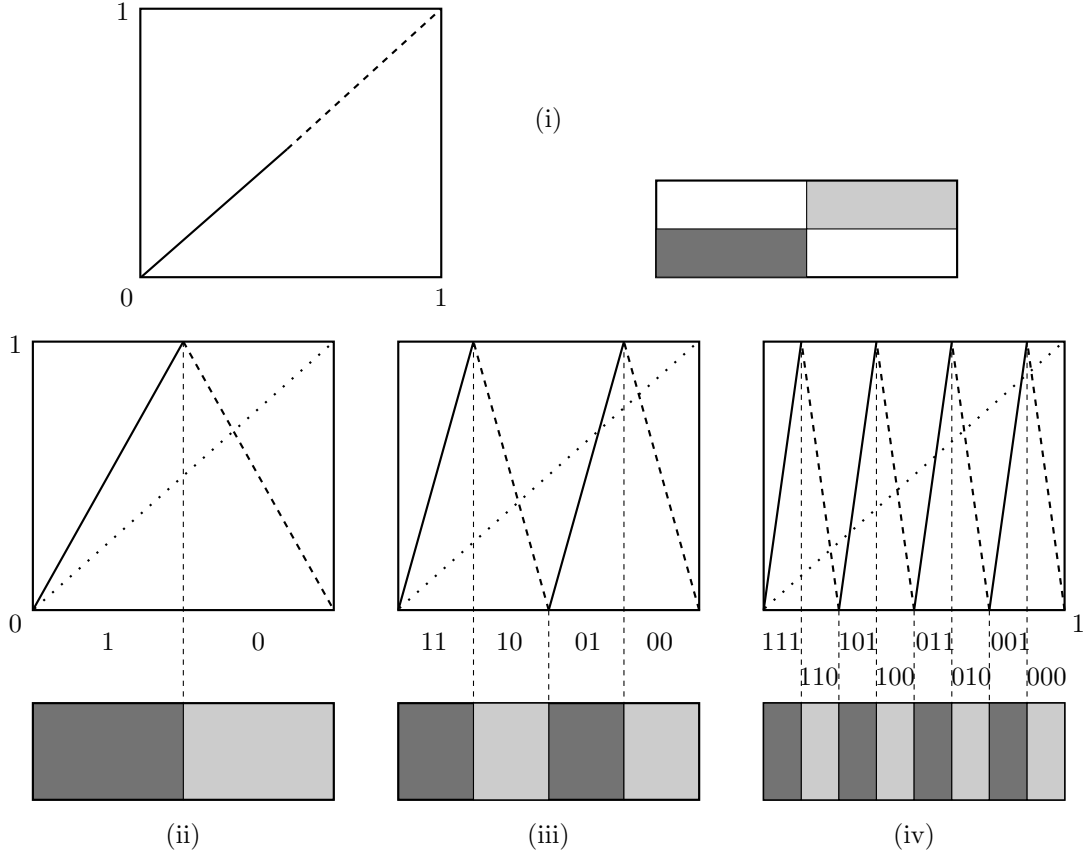}\end{center}

\caption{{\small \label{fig: dynamics}Matter-antimatter synthesis of an evolving system
$C=A\cup B$ under the tent interaction. Here $A$ is represented by the solid
line while the dashed corresponds to the anti-set $\mathfrak{{B}}$. Taking
$T_{A}>T_{B}$, $p_{A}>p_{B}$ and $\mu_{A}>\mu_{B}$, the dynamical evolution
expressed by the shaded boxes would, in the absence of backward evolution induced
by the anti-space, eventually be spread uniformly over the full domain, and
equilibrium would be characterized solely by $T_{A}$, $p_{A}$, $\mu_{A}$
from the complete annihilation of $B$. The backward evolution from the exclusion
topology of $\mathfrak{{B}}$ leads to the equilibrated stasis shown. Denoting
matter by $1$ and (the effect of) anti-matter by $0$, a progressively refined
partition of $\mathscr{D}(t)$ generated by the evolving map is indicated in
(ii), (iii) and (iv), where the partitioning sequence indicated is not that
generated by the symbolic dynamics of the map. In the real world of the figure,
matter-antimatter components are distinguished as injective branches with positive
and negative slopes, and it is seen that their directional arrows oppose each
other. And it is of course only this (increasing) non-injectivity that leads
to the interesting competing collaborative dynamics of $C$.}}
\end{figure}

In summary, then, for an interaction $f\!:C\rightarrow C$ and the bijective
map $\mathfrak{{f}}\!:C\rightarrow\mathfrak{{C}}$ of Eq. (\ref{eq: matter-anti(b)}),
the hierarchal order \begin{align*}
\textrm{Dynamics of $\mathfrak{{f}}f\!:C\rightarrow\mathfrak{{C}}$ in nature's }kitchen\,(C,\mathcal{{W}})\times(\mathfrak{{C}},\mathcal{{W}}_{\ominus})\\
\longrightarrow\textrm{Evolution of }f\textrm{ on }(C,\mathcal{{W}})^{2}\\
\longrightarrow\textrm{Experimental observables in }\mathscr{D}(f)=C\end{align*}

\noindent governed by 

$\blacktriangleright$ Basic global irreversible unidirectional evolution of
$f$ driven by ininality of topology generated on $C$ by the interaction $f$.
The function-multifunction asymmetry between $f$ and $f^{-}$ is responsible
for the unidirectionality of ininality,

$\blacktriangleright$ Induced local bi-directional dynamics of $f$ in $C^{2}$
generated by the inhibitory influence of the anti-space $(\mathfrak{{C}},\mathcal{{W}}_{\ominus})$
on $(C,\mathcal{{W}})$ that moderates the global forward evolution in $C^{2}$
to a state of dynamical balance between the competitively collaborating interactions
generated by $f$ and $f^{-}$,

\noindent define the state of equilibrated stasis schematized in Fig. \ref{fig: stasis}.
Recalling the discussion in connection with Fig. \ref{fig: GenInv1} that ininality
is an effective expression of \emph{non-bijective homeomorphicism} in which
the sequence of evolutions $(f^{n})$ become progressively bijective, according
to (\ref{eq: bijective_homeo}), on the saturated open sets of equivalence classes
and their respective images, it can be argued that \emph{the incentive towards
the resulting effective simplicity of invertibility on the definite classes
of sets associated with} $(f^{n})$ \emph{is responsible for the evolutionary
dynamics on} $C$. 

The present exposition of {}``providing a mechanical (i.e., dynamical) explanation
of why classical systems behave thermodynamically'' \citep{Callender1999},
is to be compared and contrasted with the approaches of \citet{Goldstein2004}
and \citet{Callender1999}, see also \citet{Sklar1993}. The fundamental point
of departure of our formulation lies in its non-subscription to the Newtonian
paradigm of \emph{microscopic} Hamilton's equations yielding the Liouville equation
for \emph{}density \emph{}distribution \emph{}of \emph{macroscopic} mechanical
processes; as so eloquently espoused by \citet{Baranger2000}, can the emerging
evolutionary properties of strongly nonlinear, self-organizing systems be successfully
modelled by linear (Hamiltonian) differential equations (of motion)? By employing
functional interactions as solutions to \emph{difference equations} by the technique
of graphical convergence of their iterates (in preference to linear differential
equations), we explicitly involve the past in predicting the future and are
thereby able to circumvent the issues of time reversal invariance and Poincare
recurrence that are inherently associated with the microscopic dynamics of Hamilton's
differential equations. This also enables us to avoid direct reference to statistical
and probabilistic arguments except in so far as implied by the Axiom of Choice.

\subsection{\noindent \label{sub: Complexity}An Index of Nonlinearity: Complexity }

With ininality in the cartesian space $C\times C$ serving as the engine for
the increase of evolutionary entropic disorder, we now examine how a specifically
nonlinear index can be ascribed to chaos, nonlinearity and complexity, and thereby
to serve as the benchmark for chanoxity. For this, we first recall two non-calculus
formulations of entropy that measure the complexity of dynamics of evolution
of a map $f$. 

Let $\mathcal{{A}}=\{ A_{i}\}_{i=1}^{I}$ be a disjoint partition of non-empty
subsets of a set $X$; thus ${\scriptstyle \bigcup\,}_{i=1}^{I}A_{i}=X$. The
entropy \begin{align}
S(\mathcal{{A}}) & =-\sum_{i=1}^{I}\mu(A_{i})\ln(\mu(A_{i})),\qquad\sum_{i=1}^{I}\mu(A_{i})=1\label{eq: Shannon}\end{align}
 of the partition $\mathcal{{A}}$, with $\mu(A_{i})$ some normailzed invariant
measure of the elements of the partition, quantifies the uncertainty of the
outcome of an experiment on the occurrence of any element $A_{i}$ of the partition
$\mathcal{{A}}$. A refinement $\mathcal{{B}}=\{ B_{j}\}_{j=1}^{J\geq I}$ of
the partition $\mathcal{{A}}$ is another partition such that every $B_{j}$
is a subset of some $A_{i}\in\mathcal{{A}}$, and the largest common refinement\begin{align*}
\mathcal{{A}}\bullet\mathcal{{B}} & \overset{\textrm{def}}=\{ C\!:C=A_{i}\,{\textstyle \bigcap}\, B_{j}\textrm{ for some }A_{i}\in\mathcal{{A}},\textrm{ and }B_{j}\in\mathcal{{B}}\}\end{align*}
 of $\mathcal{{A}}$ and $\mathcal{{B}}$ is the partition whose elements are
intersections of those of $\mathcal{{A}}$ and $\mathcal{{B}}$. The entropy
of $\mathcal{{A}}\bullet\mathcal{{B}}$ is given by \begin{eqnarray}
S(\mathcal{{A}}\bullet\mathcal{{B}}) & \!\!=\!\! & S(\mathcal{{A}})+S(\mathcal{{B}}\mid\mathcal{{A}})\label{eq: cond_entropy}\\
 & \!\!=\!\! & S(\mathcal{{B}})+S(\mathcal{{A}}\mid\mathcal{{B}}),\nonumber \end{eqnarray}
 where the weighted average \sublabon{equation}\begin{eqnarray}
S(\mathcal{{B}}\mid\mathcal{{A}}) & \!\!=\!\! & \sum_{i=1}^{I}P(A_{i})\, S(\mathcal{{B}}\mid A_{i})\label{eq: cond_entropy(a)}\end{eqnarray}
of the conditional entropy \begin{eqnarray}
S(\mathcal{{B}}\mid A_{i}) & \!\!=\!\! & -\sum_{j=1}^{J}P(B_{j}\mid A_{i})\,\ln(P(B_{j}\mid A_{i}))\label{eq: cond_entropy(b)}\end{eqnarray}
 of $\mathcal{{B}}$ given $A_{i}\in\mathcal{{A}}$, is a measure of the uncertainty
of $\mathcal{{B}}$ if at each trial it is known which among the events $A_{i}$
has occurred, and \begin{eqnarray}
P(B_{j}\mid A_{i}) & \!\!=\!\! & \frac{P(B_{j}\cap A_{i})}{P(A_{i})}\label{eq: cond_entropy(c)}\end{eqnarray}
\sublaboff{equation}yields the probability measure $P(B_{j}\cap A_{i})$ from
the conditional probability $P(B_{j}\mid A_{i})$ of $B_{j}$ given $A_{i}$,
with $P(A)$ the probability measure of event $A$. 

The entropy (\ref{eq: Shannon}) of the refinement $\mathcal{{A}}^{n}$, rather
than (\ref{eq: cond_entropy}), which has been used by Kolmogorov in the form
\begin{align}
h_{\textrm{KS}}(f;\mu) & \overset{\textrm{def}}=\sup_{\mathcal{{A}}_{0}}\left(\lim_{n\rightarrow\infty}\frac{1}{n}\, S(\mathcal{{A}}^{n})\right)\label{eq: Kolmogorov}\end{align}
to represent the complexity of the map as measuring the time rate of creation
of information with evolution, yields $\ln2$ for the tent transformation. Another
measure the topological entropy $h_{\textrm{T}}(f):=\sup_{\mathcal{{A}}_{0}}\lim_{n\rightarrow\infty}(\ln N_{n}(\mathcal{{A}}_{0})/n)$,
where $N_{n}(\mathcal{{A}}_{0})$ is the number of divisions of the partition
$\mathcal{{A}}^{n}$ derived from $\mathcal{{A}}_{0}$ that reduces to\begin{align}
h_{\textrm{T}}(f) & =\lim_{n\rightarrow\infty}\frac{1}{n}\ln\mathscr{I}(f^{n})\label{eq: topological}\end{align}
 in terms of the number of injective branches $\mathscr{I}(f^{n})$ of $f^{n}$
for partitions generated by piecewise monotone functions, also yields $\ln2$
for the entropy of the tent map. For the logistic map, \begin{equation}
\mathscr{I}(f^{n})=\mathscr{I}(f^{n-1})+\left\langle \{ x\!:x=f^{-(n-1)}(0.5)\}\right\rangle \label{eq: lap-number}\end{equation}
yields the number of injective branches from the solutions of \begin{align*}
{\displaystyle 0=\frac{df^{n}(x)}{dx}} & ={\displaystyle \frac{df(f^{n-1})}{df^{n-1}}\,\frac{df^{n-1}(x)}{dx}}\\
 & ={\displaystyle \frac{df(f^{n-1})}{df^{n-1}}\,\frac{df(f^{n-2})}{df^{n-2}}\cdots\frac{df(f)}{df}\,\frac{df(x)}{dx}}\end{align*}

\noindent that provide \[
\begin{array}{ccl}
 &  & \quad\,{\scriptstyle (n-1)\,\textrm{times}}\\
x & \!\!=\!\! & \overbrace{f^{-}(\cdots(f^{-}(f^{-}}(0.5)))\cdots),\qquad n=1,2,\cdots,(n-1);\end{array}\]
 here $\left\langle \{\cdots\}\right\rangle $ is the cardinality of set $\{\cdots\}$.
It should be noted that in the context of the topological entropy, $\mathscr{I}(f)$
is merely a tool for generating a partition on $\mathscr{D}(f)$ by the iterates
of $f$. 

\noindent \textbf{Examples.} (1) In a fair-die experiment, if $\mathcal{{A}}=\{\textrm{even, odd}\}$
and the refinement $\mathcal{{B}}=\{ j\}_{j=1}^{6}$ is the set of the six faces
of the die, then for $i=1,2${\renewcommand{\arraystretch}{1.5}\[
\begin{array}{ccc}
P(B_{j}\mid A_{i}) & \!\!=\!\! & \left\{ \begin{array}{cll}
{\displaystyle \frac{1}{3}}, &  & j\in A_{i}\\
0, &  & j\notin A_{i},\end{array}\right.\end{array}\]
and $S(\mathcal{{B}}\mid A_{1})=\ln3=S(\mathcal{{B}}\mid A_{2})$ according
to Eq. (\ref{eq: cond_entropy(b)}). Hence the conditional entropy of $\mathcal{{B}}$
given $\mathcal{{A}}$, using $P(A_{1})=0.5=P(A_{2})$, is $S(\mathcal{{B}}\mid\mathcal{{A}})=\ln3$
by (\ref{eq: cond_entropy(a)}), and\begin{align*}
S(\mathcal{{A}}\bullet\mathcal{{B}}) & =S(\mathcal{{A}})+S(\mathcal{{B}\mid{A}})\\
 & =\ln6,\end{align*}
 \[
\begin{array}{ccc}
P(B_{j}\textrm{ }{\textstyle \bigcap}\textrm{ }A_{i}) & \!\!=\!\! & \left\{ \begin{array}{cll}
{\displaystyle \frac{1}{6}}, &  & j\in A_{i}\\
0, &  & j\notin A_{i}.\end{array}\right.\end{array}\]
}If we have access only to partition $\mathcal{{B}}$ and not to $\mathcal{{A}}$,
then $S(\mathcal{{B}})=\ln6$ is the amount of information gained about the
partition $\mathcal{{B}}$ when we are told which face showed up in a rolling
of the die; if on the other hand the only partition available is $\mathcal{{A}}$,
then $S(\mathcal{{A}})=\ln2$ measures the information gained about $\mathcal{{A}}$
on the knowledge of the appearance of an even or odd face. 

(2) The dynamical evolution of Fig. \ref{fig: dynamics} is a case in point
of conditional probability and conditional entropy. Here the refinements of
basic partition $\mathcal{{A}}_{0}=\{\textrm{matter, antimatter}\}=\{ A_{01},A_{00}\}$
generated by the inverses of the tent map, interpreted as representing matter-antimatter
dynamics, are denoted as $\mathcal{{A}}_{n}=\{ t^{-n}(A_{0i})\}_{0,1}$ for
$n=1,2,\cdots$ to yield the largest common refinements \begin{equation}
\mathcal{{A}}^{n}=\mathcal{{A}}_{0}\bullet\mathcal{{A}}_{1}\bullet\mathcal{{A}}_{2}\bullet\cdots\bullet\mathcal{{A}}_{n},\qquad n\in\mathbb{{N}}.\label{eq: refinement_n}\end{equation}
 With reference to Fig. \ref{fig: dynamics}, these refinements are symbolically
denoted by $\{1,0\}\rightarrow\{11,10,01,00\}\rightarrow\{111,110,101,100,011,010,001,000\}\rightarrow\cdots$,
and $\mathcal{{A}}^{n}=\mathcal{{A}}_{n}$. Taking the measure of the elements
of a partition to be its euclidean length, gives {\renewcommand{\arraystretch}{1.4}\[
\begin{array}{ccc}
P(A_{nj}\mid A_{0i}) & \!\!=\!\! & \left\{ \begin{array}{lll}
{\displaystyle \frac{1}{2^{n-1}}}, &  & j\in A_{0i}\\
0, &  & j\notin A_{0i},\end{array}\right.\end{array},\]
}$S(\mathcal{{A}}_{n}\mid A_{0i})=(n-1)\ln2$, $i=0,1$, (Eq. \ref{eq: cond_entropy(b)}),
$S(\mathcal{{A}}_{n}\mid A_{0})=(n-1)\ln2$, and finally $S(\mathcal{{A}}_{n}\bullet\mathcal{{A}}_{0})=n\ln2.$
In case the initial partition $\mathcal{{A}}_{0}$ is taken to be the whole
of $\mathscr{D}(t)$, then Eq. (\ref{eq: Shannon}) gives directly $S(\mathcal{{A}}_{n})=n\ln2$.
\begin{figure}[htbp]
\noindent \begin{center}\input{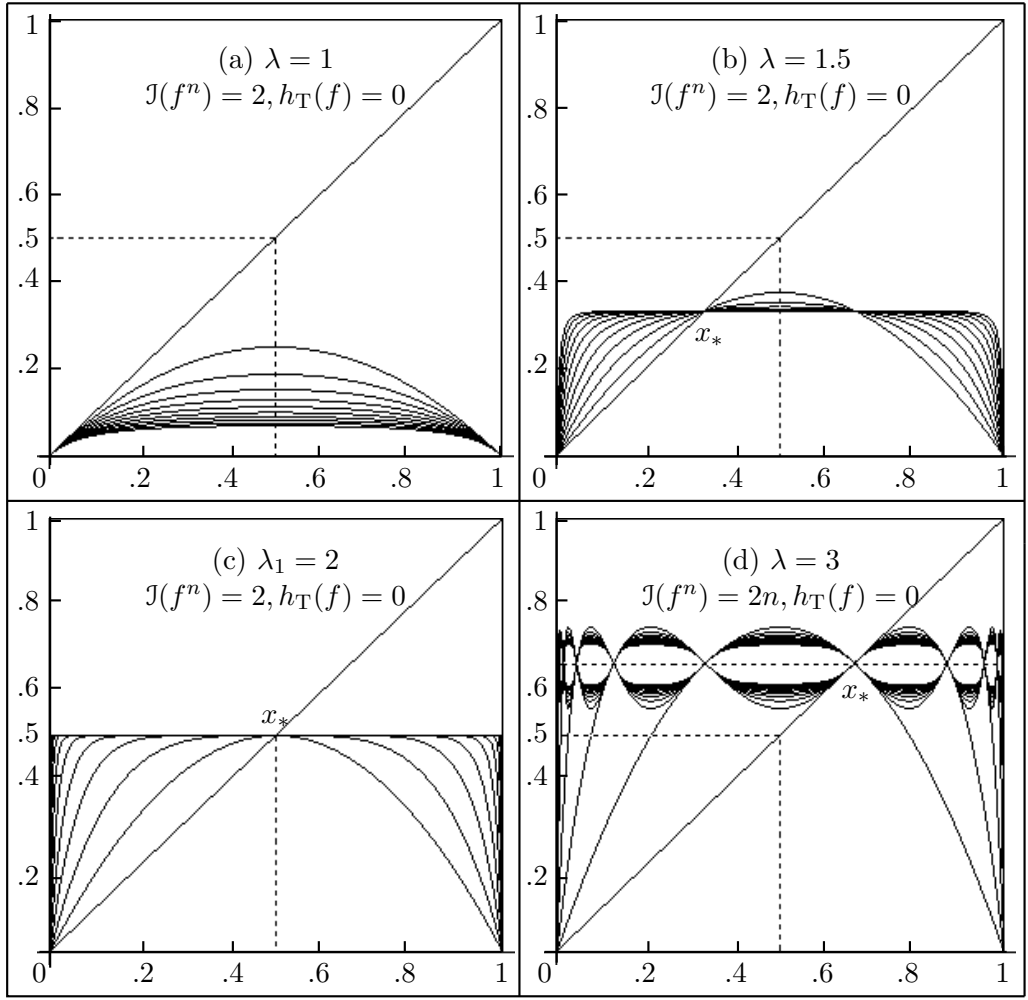}\end{center}

\caption{{\small \label{fig: cmplx_stable}Non-life dynamics of the first 10 iterates
of the logistic map $f_{\lambda}=\lambda x(1-x)$ generated by its only stable
fixed point $x_{*}=(\lambda-1)/\lambda$. Although the partition induced on
$X=[0,1]$ by the evolving map in (d) is refined with time, the} \emph{\small stability
of the fixed point} {\small $x_{*}=0.6429$ prevents the dynamics from acquiring
any meaningful evolutionary significance with the multifunctional graphical
limit, indicated by the broken line, being of the same type as in (b) and (c):
as will be evident in the following,} \emph{\small instability of the fixed
point is necessary for the evolution of a meaningful complex life}{\small .
$\lambda_{1}=2$ of (c) --- obtained by solving the equation $f_{\lambda}(0.5)=0.5$---
is special because its super-stable} \emph{\small }{\small fixed point $x=0.5$
is the only point in $\mathscr{D}(f)$ at which $f$ is injective and therefore
well-posed by this criterion.}}
\end{figure}

\smallskip{}
(3) Logistic map $f_{\lambda}(x)=\lambda x(1-x)$, \citet{Nagashima1999}.

(i) $0<\lambda\leq3$, Fig. \ref{fig: cmplx_stable}, can be subdivided into
two categories. In the first, for $0<\lambda<\lambda_{1}=2$, $\mathscr{I}(f_{\lambda}^{n})=2$
gives $h_{\textrm{T}}(f_{\lambda})=0$. This is illustrated in Fig. \ref{fig: cmplx_stable}(a),
(b), and (c) which show how the number of subsets generated on $X$ by the increasing
iterates of the map tend from 2 to 1 in the first case and to the set $\{\left\{ 0\right\} ,(0,1),\left\{ 1\right\} \}$
for the other two. The figure demonstrates that while in (a) the dynamics eventually
collapses and dies out, the other two cases are equally uneventful in the sense
that the converged multifunctional limits --- of $(0,[0,1/2])\cup((0,1),1/2)\cup(1,[0,1/2])\}$
in figure (c), for example --- are as much passive and displays no real {}``life'';
this is quantified by the constancy of the lap number and the corresponding
topological entropy $h_{\textrm{T}}(f)=0${\large .} 

Although the oscillations in (d) for $\lambda=3$ show more apparent {}``life''
than the other cases, the iterates converge graphically to the tame $\{(0,[0,2/3])\cup([0,1],2/3)\cup(1,[0,2/3])\}$
indicated by the broken line and the topological entropy is again 0. \sublabon {figure}%
\begin{figure}[htbp]
\begin{center}\input{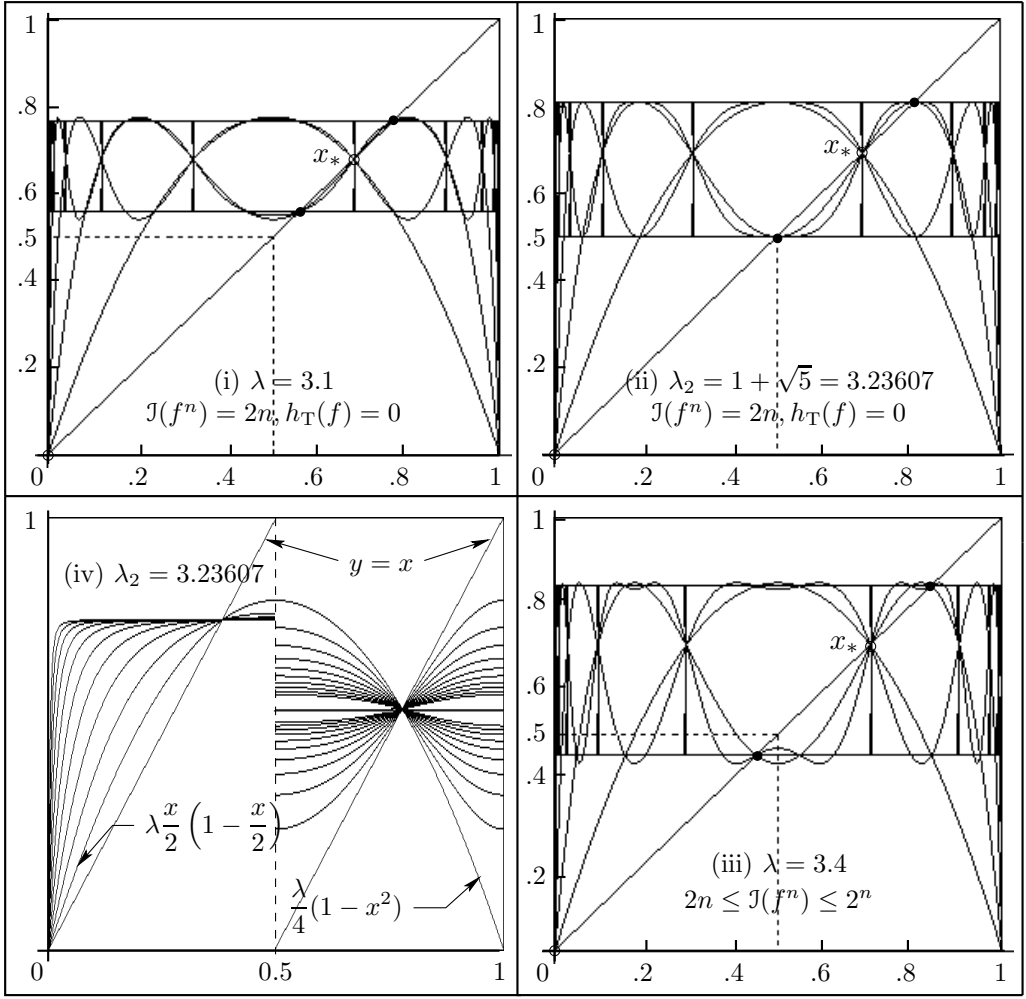}\end{center}

\caption{\noindent {\small \label{fig: cmplx_2cycle}Dynamics of stable 2-cycle of the
logistic map, where each panel displays the first four iterates superposed on
the graphically converged multifunction represented by iterates 1001 and 1002.
The unstable fixed point $x_{*}$ of $f$ is directly linked to its stable partners
$a$ and $b$ of $f^{2}$ that report back to their master $x_{*}$. Compared
to case (i) where the relative simplicity of the instability of $x_{*}$ allows
its stable partners to behave monotonically as in Fig. \ref{fig: cmplx_stable}(b),
in (iii) the instability is strong enough to induce the oscillatory mode of
convergence of Fig. \ref{fig: cmplx_stable}(iv). Case (ii) of the super-stable
cycle for $\lambda_{2}=1+\sqrt{5}$ --- obtained by solving the equation $f_{\lambda}^{2}(0.5)=0.5$
--- reflecting well-posedness of $f$ at $x=0.5$ represents, as in Fig \ref{fig: cmplx_stable}(iii),
a mean of the relative simplicity of (a) and the complex instability of (iii)
that grows with increasing $\lambda$ due to the fact that $\lambda>\lambda_{2}$
ensures $f_{\lambda}^{2}(0.5)=f_{\lambda}(\lambda/4)<0.5$. Notice that in all
the three cases, $\left\langle \{ x\!:f_{\lambda}^{n}(x)=0.5\}\right\rangle =2$
for all $n$. Panel (iv), in this and the following two subdiagrams, illustrates
how the individual parts, acting independently on their own in the reductionist
framework not in competitive collaboration, leads to an entirely different simple,
non-complex, dynamics.}}
\end{figure}

\begin{figure}[htbp]
\noindent \begin{center}\input{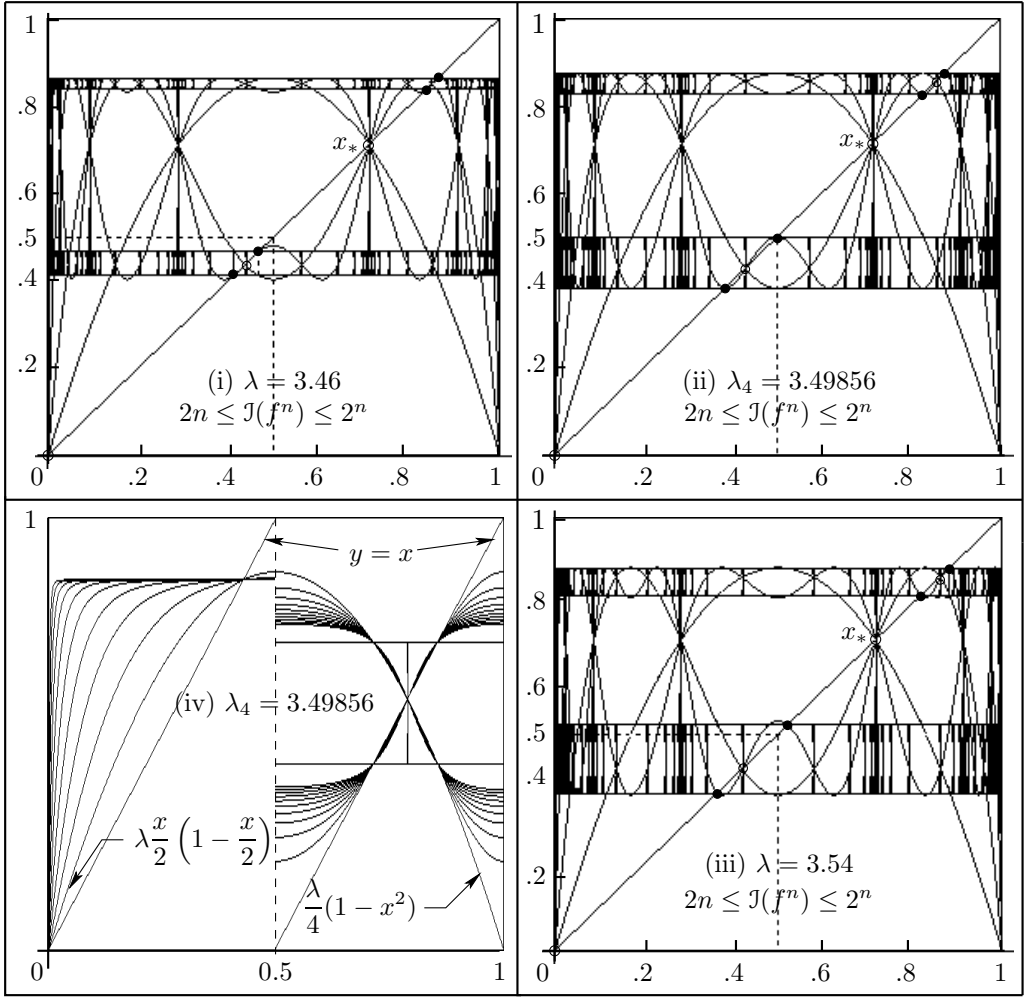}\end{center}

\caption{\noindent \label{fig: cmplx_4cycle}{\small Dynamics of stable 4-cycle of the
logistic map, where each panel displays the first four iterates superposed on
the graphically {}``converged'' multifunction represented by iterates 1001-1004.
The unstable basic fixed point $x_{*}$ due to $f$ is now linked to its} \emph{\small un}{\small stable
partners $\{ a\textrm{\} and }\{ b\}$ denoted by open circles arising from
$f^{2}$, who report back to the overall controller $x_{*}$ the information
they receive from their respective stable committees $\{ c,e\}$ and $\{ d,f\}$.
Compared to the 2-cycle of Fig. \ref{fig: cmplx_2cycle}, the instability of
principal $x_{*}$ is now serious enough to require sharing of the responsibility
by $\{ a\}$ and $\{ b\}$ who are further constrained to delegate authority
to the subcommittees mentioned above. Case (ii) of the super-stable cycle for
$\lambda_{4}=3.49856$ --- obtained by solving $f_{\lambda}^{4}(0.5)=0.5$ ---
reflecting well-posedness of $f$ at $x=0.5$ represents as before the mean
of the relative simplicity of (i) and the large instability of (iii).}}
\end{figure}

(ii) $3<\lambda\leq4$. As in (i), $h_{\textrm{T}}(f_{\lambda})=0$ whenever
$\mathscr{I}(f_{\lambda})\leq2n$ which occurs, according to Fig. \ref{fig: cmplx_2cycle},
for $\lambda\leq\lambda_{2}=1+\sqrt{5}=3.23607$; here $\lambda_{n}$ is the
$\lambda$ value at which a super-stable $n$-cycle appears. The super-stable
$\lambda$ for which $x=0.5$ is fixed with respect to $f^{n}$, $n=2^{m},\, m=0,1,2,\cdots$,
are of special significance as this is the only point in $X$ at which $f$
is injective; this leads to a simplification of the dynamics of the map that
can be verified by comparing the plots in Figs. \ref{fig: cmplx_stable}, \ref{fig: cmplx_2cycle},
and \ref{fig: cmplx_4cycle}. These super-cycles possess the great simplifying
property that the stable horizontal parts of the graphically converged mulitfunction
are actually tangential to all the turning points of every iterate of $f$.
The immediate consequence of this is that for a given $3<\lambda<\lambda_{*}=3.5699456$,
with $\lambda_{*}$ the value at the {}``edge of chaos'', the dynamics of
$f$ attains a state of basic evolutionary stability after only the first $\{2^{m}\}_{m\in\mathbb{{N}}}$
time steps in the sense that no new spatial structures \emph{emerge} after this
period, any further temporal evolution being fully utilized in spatially \emph{self}-\emph{organizing}
this basic structure throughout the system by the generation of equivalence
classes of the initial $2^{m}$ time steps. 

When $\lambda>\lambda_{2}$ as in Figs. \ref{fig: cmplx_2cycle} and \ref{fig: cmplx_4cycle}
the number of injective branches satisfy $2n\leq\mathscr{I}(f_{\lambda}^{n})\leq2^{n}$
and it is easily appreciated, from Eq. (\ref{eq: lap-number}), the difficulty
in actually calculating these numbers for large $n$. For $\lambda=4$, however
$\mathscr{I}(f_{4}^{n})=2^{n}$ and the topological entropy reduces to the simple
$h(f_{4})=\ln2$; $h_{\textrm{T}}(f)>0$ is sufficient condition for $f_{\lambda}$
to be chaotic. The tent map behaves similarly and has an identical topological
entropy, see Fig. \ref{fig: Complex-Chaos}. %

The difficulty in evaluating $\mathscr{I}(f^{n})$ for large values of $n$
and the open question of the utility of the number of injective branches of
a map in actually measuring the complex dynamics of the nonlinear evolution
of the logistic map suggests the significance of the role of \emph{evolution}
\emph{of} \emph{the graphs of the iterates of} $f_{\lambda}$ in defining the
nonlinear dynamics of natural processes. It is also implied that the dynamics
can be simulated through \emph{the} \emph{partitions induced on} $\mathscr{D}(f)$
\emph{by the evolving map} as described by graphical convergence of the functions
in accordance with our philosophy that the dynamics on $C$ derives from the
evolution of $f$ in $C^{2}$ as observed in $\mathscr{D}(f)$. The following
subsection carries out this line of reasoning, to be compared with that embodied
in Eqs. (\ref{eq: Kolmogorov}) and (\ref{eq: topological}), to define a new
index of chaos, nonlinearity and complexity, that of \emph{chanoxity}.

\subsubsection{ChaNoXity: A Measure of Chaos, Nonlinearity and Complexity}

\begin{figure}[htbp]
\begin{center}\input{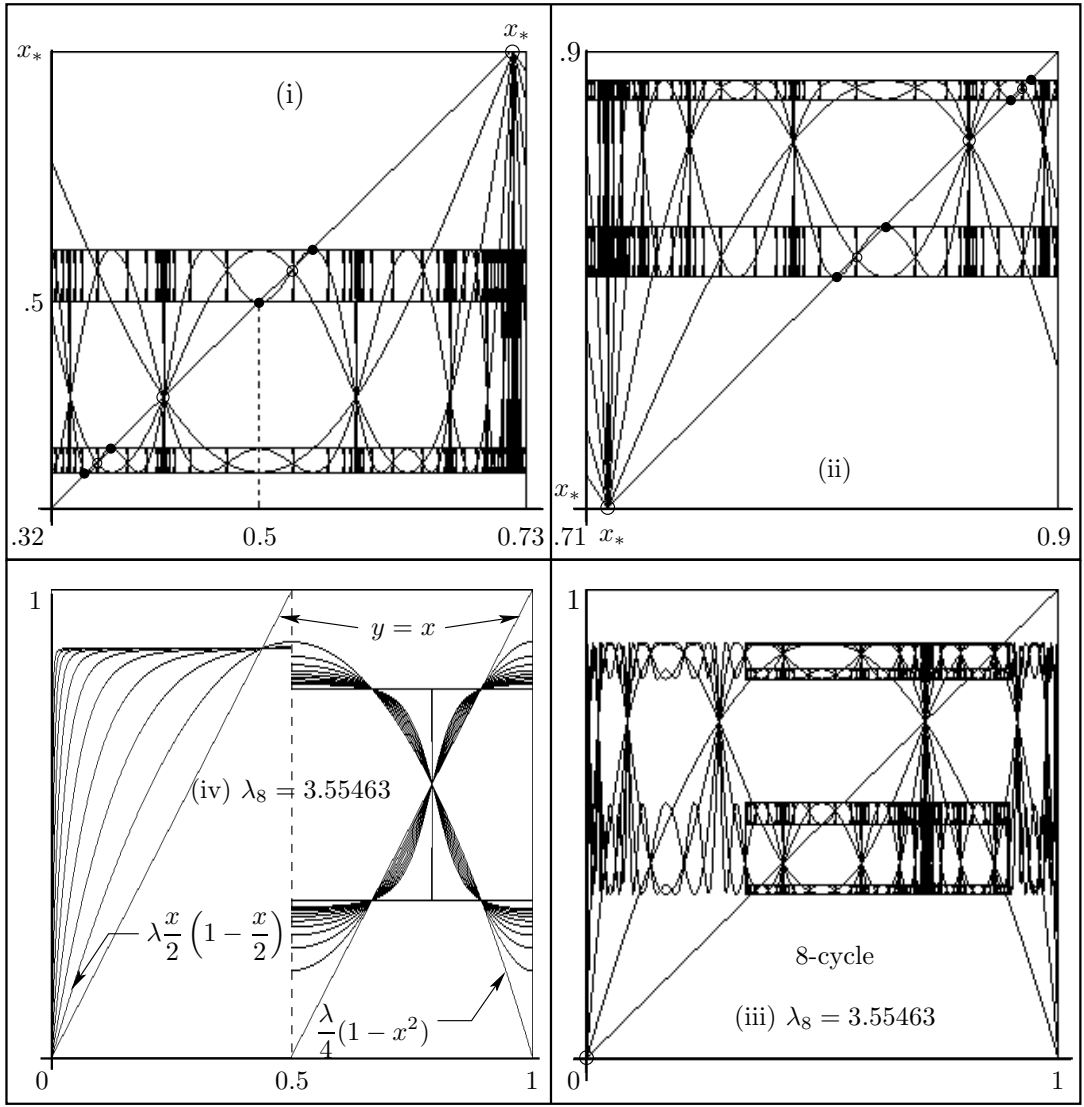}\end{center}

\caption{\noindent {\small \label{fig: cmplx_8cycle}Blown-up view of the} \emph{\small stable}
{\small 8-cycle graphically illustrates evolutionary dynamics arising from the
logistic interaction. The $2^{3}$} \emph{\small unstable} {\small fixed points
marked by open circles interact among themselves as indicated to generate the
stable periodic cycle providing thereby a vivid demonstration of competitive
collaboration between world-antiworld effects. Display (iii) of the partially
superimposed limit graphs 1001-1008 --- that remain invariant with further temporal
evolution --- demonstrates that while nothing new emerges} \emph{\small }{\small after
the first 8 time steps, further evolution consolidates the associated changes
throughout the system in the form of self-generated equivalence classes that
act as the inhibitors toward a state of eventual stasis.}}
\end{figure}

\noindent The blown-up view of the \emph{stable} 8-cycle, Fig. \ref{fig: cmplx_8cycle},
of the logistic map graphically illustrates evolutionary dynamics arising from
this interaction. The $2^{3}$ \emph{unstable} fixed points marked by open circles
interact among themselves as implied in the figure to generate the stable periodic
cycle, providing thereby a vivid illustration of competitive collaboration between
world-antiworld effects. The \emph{self-organized collaboration} is due to the
emergent irreversible urge toward bijective simplicity of ininality accompanying
increasing $\lambda$ as manifest in the Second Law increase of entropy; this
is inhibited locally for a fixed $\lambda$ by an opposing \emph{competitive
anti-effect} that eventually leads to the stable periodic orbit. This local
inhibitory restraining effect appears in the figure as the opposing change of
slope associated with each of the unstable fixed points except the first at
$x=0$ which must now be paired with its equivalent image at $x=1$. Display
(c) of the partially superimposed limit graphs 1001-1008 --- that remain invariant
with further temporal evolution --- on the first 8 iterates illustrate that
while nothing new emerges \emph{}beyond this initial period, \emph{further temporal
evolution propagates the associated changes throughout the system as self-generated
equivalence classes acting as inhibitors that restrain the system to a state
of local} (that is spatial, for the given $\lambda$) \emph{periodic stasis}.
As compared to Fig. \ref{fig: dynamics} for the tent interaction, this manifestation
of antieffects in the logistic for $\lambda<\lambda_{*}=3.5699456$ has a profound
feature that deserves attention: while in the former the antimatter branch belongs
to distinct fixed points of equivalence classes, in the later matter-antimatter
competitive-collaboration is associated with each of the $2^{N}$ generating
fixed point branches possessing bi-directional characteristics with the inhibition
of antimatter actually generating the equivalence class. In the observable real
world of $\mathscr{D}(f)$, this has the interesting consequence that whereas
the tent interaction generates matter-antimatter intermingling of disjoint components
to produce the homogenization of Fig. \ref{fig: dynamics}, for the logistic
the resulting behaviour \emph{}is a consequence of a deeper interplay of the
opposing forces leading to a higher level of complexity than can be achieved
by the tent interaction. This distinction reflects in the interaction pair $(f,\mathfrak{{f}})$
that can be represented as \begin{equation}
x\mapsto2x\overset{\mathfrak{{tent}}}\longmapsto\left\{ \begin{array}{ll}
2x, & \textrm{if }0\leq x<0.5\\
2(1-x), & \textrm{if }0.5\leq x\leq1\end{array}\right.,\qquad x\mapsto2x\overset{\mathfrak{{logistic}}}\longmapsto4x(1-x),\label{eq: (f, mathfrak{f}))}\end{equation}

\noindent which leads to the 

\smallskip{}
\noindent \textbf{Definition. Complex System, Complexity.} The couple $((X,\mathcal{{U}}),f)$
of a topological space $(X,\mathcal{{U}})$ and an interaction $f$ on it is
a \emph{complex system} if 

(CS1) The \emph{algebraic} \emph{structure} of $X$ consists of a family of
progressively refined disjoint partitions of non-empty subsets induced by the
iterates of $f$. Of these, only a \emph{coarsest} \emph{finite} \emph{number
determines the character of the system,} with successive refinements erecting
on this defining foundation the structure of the system. 

(CS2) The \emph{topology} $\mathcal{{U}}$ of $X$ is generated \emph{through
a process of} \emph{competitive} \emph{collaboration} \emph{among the hierarchy
of partitions} in the sense that the subbasis of $\mathcal{{U}}$ at any level
of refinement is the union of the open sets of its immediate coarser partition
and that generated independently by the partition under consideration; here
all open sets are the saturated sets of equivalence classes under the iterates
of the interaction. 

The \emph{complexity} of a system is a measure of the interaction between the
different levels of partitions that are generated on $\mathscr{D}(f)$ under
the induced topology on $X$. %
\sublaboff {figure}

Thus for example in Fig. \ref{fig: cmplx_2cycle}(ii) of the stable 2-cycle,
the defining character is established by just the first 2 time steps which is
then propagated throughout the system by the increasing ill-posednesss of the
future iterates, thereby establishing the global structure as seen in the diagram.
The open sets of $\mathscr{D}(f)$ are the projections of the boxes onto the
$x$-axis, with their boundary being represented by the members of the equivalence
class $[x_{*}]$ of the unstable fixed point $x_{*}$. With increasing $\lambda$
the complexity of the dynamics increases as revealed by the succeeding plots
of 4- and 8-cycles; this allows us to define the \emph{chanoxity index} of the
interaction to be the \emph{constant} $\nu$ that satisfies \sublabon {equation}\begin{equation}
f(x)=x^{1-\nu},\qquad\forall x\in\mathscr{D}(f).\label{eq: Index_Nu(a)}\end{equation}
If $\left\langle f(x)\right\rangle $ and $\left\langle x\right\rangle $ are
the measures that make Eq. (\ref{eq: Index_Nu(a)}) possible, then in \begin{equation}
\nu\overset{\textrm{def}}=1-\frac{\ln\left\langle f(x)\right\rangle }{\ln\left\langle x\right\rangle }\label{eq: Index_Nu(b)}\end{equation}
\sublaboff {equation} we adopt the criteria that 

(a) $\left\langle x\right\rangle $ is the number of basic unstable fixed points
of $f$ for any $\lambda$ that is responsible for \emph{emergence}. Thus for
$1<\lambda\leq3$ there is just one unstable fixed point at $x=0$, which is
then followed by the familiar sequence of $2^{N}$ fixed points until $\lambda=\lambda_{*}$
when this number is infinite. 

(b) For $f(x)$ the estimate\begin{align*}
\left\langle f(x)\right\rangle  & =2f_{1}+\sum_{j=1}^{N}\sum_{i=1}^{2^{N-j}}f_{i,2^{N-j}},\quad N=1,2,\cdots,\end{align*}
where $f_{i}=f^{i}(0.5)$ and $f_{i,j}=|f^{i}(0.5)-f^{j}(0.5)|$, leads to the
measure of the chanoxity index which for understandable reasons we call the
dimensional chanoxity of f\_\{\textbackslash{}lambda\} .\begin{align}
\nu_{N} & =1-\frac{1}{N\ln2}\,\ln\left[2f_{1}+\sum_{j=1}^{N}\sum_{i=1}^{2^{N-j}}f_{i,2^{N-j}}\right],\label{eq: chanoxity_nu}\end{align}
which which for understandable reasons we call the \emph{dimensional chanoxity
of} $f_{\lambda}$.for understandable reasons we call the \emph{dimensional
chanoxity of} $f_{\lambda}$.%
\footnote{\label{foot: fractal}Recall that the fractal dimension of an object is formally
defined very similarly: \[
D=\frac{\ln(\textrm{\# self-similar pieces into which the object can be decomposed})}{\ln(\textrm{magnification factor that restores each piece to the original})}.\]
} In the calculations reported here, $\lambda$ is taken to correspond to the
respective superstable periodic cycle, where we note from Figs. \ref{fig: cmplx_stable},
\ref{fig: cmplx_2cycle} and \ref{fig: cmplx_4cycle} that the corresponding
super-stable dynamics faithfully reproduces all the features of emergence and
self organization of that stable $2^{N}$ cycle class. \sublabon{table} %
\begin{table}[htbp]
\noindent \begin{center}{\renewcommand{\arraystretch}{1.4}\begin{tabular}{|c|c|c|c|c|}
\hline 
$\lambda$&
$N$&
$f^{2N}(0.5)$&
$\left\langle f(0.5)\right\rangle $&
$\nu_{N}$\tabularnewline
\hline
\hline 
2.0000000&
$-$&
0.50000&
1.000000&
0.000000\tabularnewline
\hline 
3.2360680&
1&
0.50000&
1.927051&
0.053605\tabularnewline
\hline 
3.4985600&
2&
0.50000&
2.404122&
0.367245\tabularnewline
\hline 
3.5546300&
3&
0.50001&
2.680845&
0.525771\tabularnewline
\hline 
3.5666700&
4&
0.50000&
2.842181&
0.623250\tabularnewline
\hline 
3.5692401&
5&
0.50001&
2.935103&
0.689318\tabularnewline
\hline 
3.5697999&
6&
0.50004&
2.989741&
0.736663\tabularnewline
\hline 
3.5699124&
7&
0.50001&
3.019392&
0.772249\tabularnewline
\hline 
3.5699439&
8&
0.50014&
3.043283&
0.799296\tabularnewline
\hline 
3.5699446&
9&
0.49994&
3.050981&
0.821192\tabularnewline
\hline 
3.5699451&
10&
0.50016&
3.059756&
0.838658\tabularnewline
\hline
$\downarrow$&
$\downarrow$&
$\vdots$&
$\downarrow$&
$\downarrow$?\tabularnewline
\hline 
3.5699456&
$\infty$&
$-$&
3.??????&
1.000000\tabularnewline
\hline
\end{tabular}}\end{center}

\caption{{\small \label{tab: Index_Nu(a)} In the passage to full chaoticity, the system
becomes increasingly complex and nonlinear (remember: chaos is maximal nonlinearity)
such that at the edge of chaos $\lambda=\lambda_{*}=3.5699456$, the system
is completely complex and chaotic with $\nu=1$. For $1<\lambda\leq3$ with
no generated instability of which $\lambda=2$ is representative, $\nu=1-\ln(1/2+1/2)/\ln1$.}}
\end{table}

The numerical results of Table \ref{tab: Index_Nu(a)} suggest that \[
\lim_{N\rightarrow\infty}\nu_{N}=1\]
at the {}``edge of chaos'' $\lambda=\lambda_{*}=3.5699456$. Since $\nu=0$
gives the simplest linear relation for $f$, values of $1$ and $-\infty$ for
the index indicate the largest non-linearity and complexity so that the logistic
interaction is maximally complex at the transition to the fully chaotic region.
For this range of values $3\leq\lambda\leq\lambda_{*}$, the associated increasing
energy input to the system is fully utilized in enhancing its complexity through
increasing structural emergence with the accompanying self-organization transmitting
this emerging behaviour throughout the system as enumerated earlier. %
\begin{table}[htbp]
\noindent \begin{center}{\renewcommand{\arraystretch}{1.2}\begin{tabular}{|c|c|c|c|c|c|c|c|c|}
\hline 
\multirow{2}{*}{$\lambda$}&
&
\multicolumn{7}{c|}{$N$} \\cline{3-9}&
&
&
&
&
&
\tabularnewline
&
&
12&
14&
16&
18&
20&
$\rightarrow$&
$\infty$\tabularnewline
\hline
\hline 
\multirow{2}{*}{3.57}&
$\left\langle f(0.5)\right\rangle $&
4.387099&
6.626634&
13.76167&
40.47870&
145.5237&
&
\tabularnewline
&
$\nu_{N}$&
0.822228&
0.780513&
0.763589&
0.703384&
0.640744&
$\overset{?}\rightarrow$&
0.0000\tabularnewline
\hline 
\multirow{2}{*}{3.6}&
$\left\langle f(0.5)\right\rangle $&
290.3677&
1171.071&
4555.594&
17900.68&
73980.36&
&
\tabularnewline
&
$\nu_{N}$&
0.318189&
0.271885&
0.241411&
0.215127&
0.191258&
$\overset{?}\rightarrow$&
0.0000\tabularnewline
\hline 
\multirow{2}{*}{3.7}&
$\left\langle f(0.5)\right\rangle $&
950.8090&
3828.215&
14796.66&
61351.94&
236962.9&
&
\tabularnewline
&
$\nu_{N}$&
0.175582&
0.149825&
0.134188&
0.116399&
0.107285&
$\overset{?}\rightarrow$&
0.0000\tabularnewline
\hline 
\multirow{2}{*}{3.8}&
$\left\langle f(0.5)\right\rangle $&
1162.450&
4612.551&
18565.80&
74061.57&
295511.7&
&
\tabularnewline
&
$\nu_{N}$&
0.151421&
0.130618&
0.113728&
0.101309&
0..091357&
$\overset{?}\rightarrow$&
0.0000\tabularnewline
\hline 
\multirow{2}{*}{3.9}&
$\left\langle f(0.5)\right\rangle $&
1390.724&
5703.793&
22384.97&
90580.02&
359594.0&
&
\tabularnewline
&
$\nu_{N}$&
0.129865&
0.108735&
0.096860&
0.085172&
0.077200&
$\overset{?}\rightarrow$&
0.0000\tabularnewline
\hline
$\downarrow$&
$\vdots$&
$\downarrow$?&
$\downarrow$?&
$\downarrow$?&
$\downarrow$?&
$\downarrow$?&
$\searrow$?&
\tabularnewline
\hline 
4.0&
$\nu_{N}$&
0.0000&
0.0000&
0.0000&
0.0000&
0.0000&
&
0.0000\tabularnewline
\hline
\end{tabular}}\end{center}

\caption{\label{tab: Index_Nu(b)}{\small Illustrates how the fully chaotic region of
$\lambda_{*}<\lambda\leq4$ is effectively {}``linear'' with self-organization
and emergence implying each other. The jump discontinuity in $\nu$ at the edge
of chaos $\lambda_{*}$ reflects a qualitative change in the dynamics, with
all the gainfully employed energy input for $\lambda\leq\lambda_{*}$ employed
in the generation of the complex internal structures of the system being fully
utilized in generating the emerging characteristics without any self-organization
as $\lambda\rightarrow4$ .}}
\end{table}
\sublaboff{table}

\begin{figure}[htbp]
\noindent \begin{center}\input{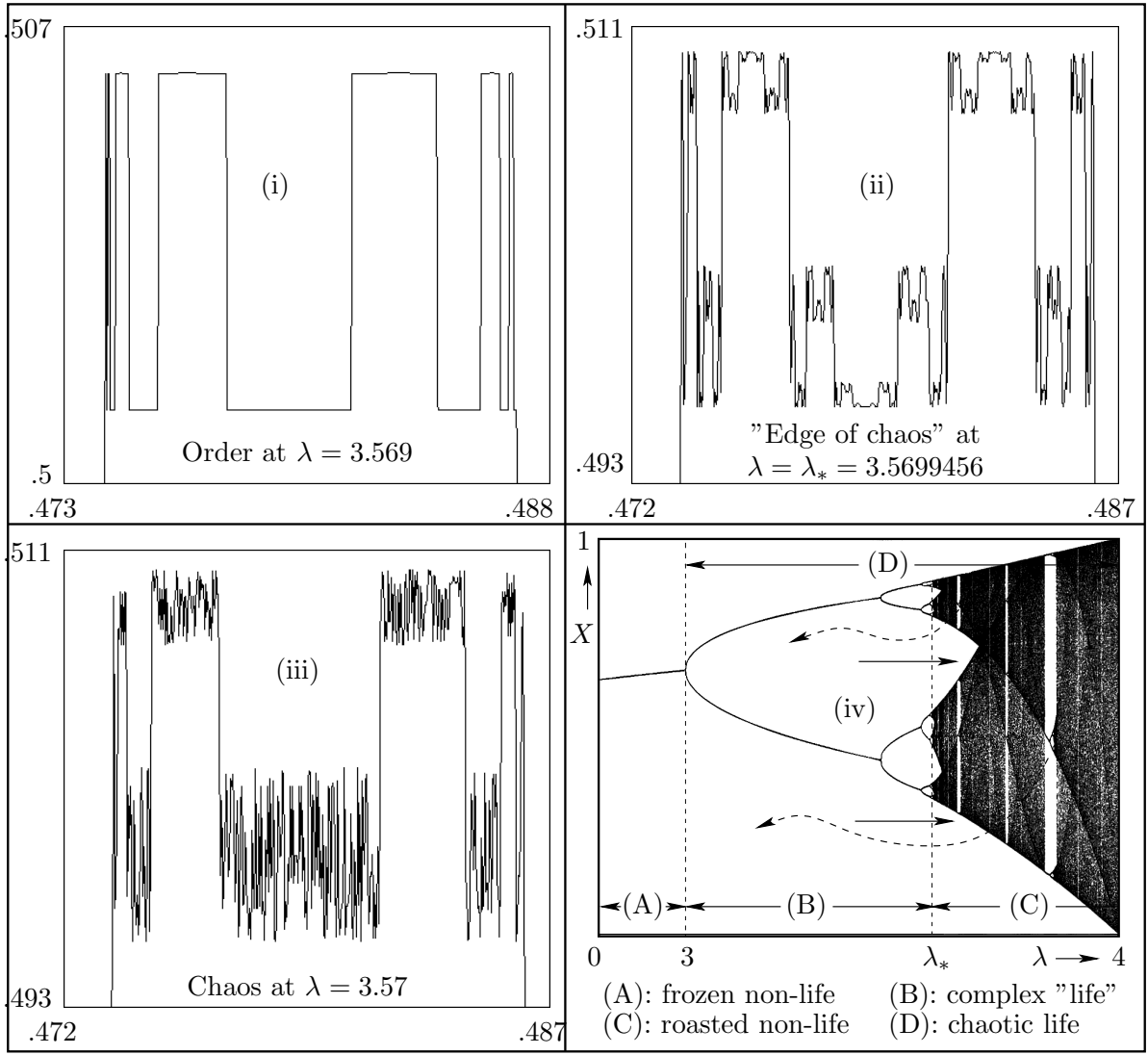}\end{center}

\caption{\label{fig: Complex-Chaos}{\small In contrast with the relatively tame (i)
and (ii), panel (iii) illustrates the property of fully chaotic maximal ill-posedness
and instability. Here $\mathscr{I}(f^{n})=2^{n}$, and the topological entropy
$h_{\textrm{T}}(f)=\lim_{n\rightarrow\infty}\ln(\mathscr{I}(f^{n}))/n=\ln2$.}
{\footnotesize In (iv) we illustrate the evolution of natural processes under
the logistic interaction as summarized in panels (i)$-$(iii) and in Eq. (\ref{eq: Complex-Chaos}).
The dashed arrows indicate Sarkovskii stabilization of the full-arrowed, entropic,
ininal drive towards a state of superheated, chaotic, non-life illusory simplicity
when the non-trivial fixed point no longer determines the fate of the evolutionary
dynamics of the system, thereby establishing a} \emph{\footnotesize chaotically
complex} {\footnotesize state of dynamical equilibrium. A typical example of
life-defining complex system (B) is the parliamentary system of governance with
the speaker of the House acting as the supreme non-trivial fixed authority of
the constitutional interaction between the ruling party and the opposition,
while the Iraq war and its aftermath offers a versatile model of chaotic complexity
(D). }}
\end{figure}

What happens for $\lambda>\lambda_{*}$ in the fully chaotic region where emergence
persists for all times $N\rightarrow\infty$ with no self-organization, is shown
in Table \ref{tab: Index_Nu(b)} which indicates that on crosssing the chaotic
edge, the system abruptly transforms \emph{}to a state of \emph{effective linear
simplicity} that can be interpreted to result from the drive toward ininality
and effective bijectivity on saturated sets and on the image component space
of $f$. This jump discontinuity in $\nu$ demarcates order from chaos, linearity
from (extreme) nonlinearity, and simplicity from complexity. This non-organizing
region $\lambda>\lambda_{*}$ of deceptive non-life simplicity characterized
by dissipation and irreversible {}``frictional losses'', is to be compared
and contrasted with the nonlinearly complex region $3<\lambda\leq\lambda_{*}$
where irreversibility generates self-organizing, useful changes in the internal
structure of the system in order to attain the levels of complexity needed in
the evolutionary process. While the state of eventual evolutionary stasis appears
in $3<\lambda\leq\lambda_{*}$, the relative linear simplicity of $\lambda>\lambda_{*}$
arising from the competitive dissipatory losses characteristic of this region
conceals the resulting self-organizing thrust on $3<\lambda\leq\lambda_{*}$
of the higher periodic windows of this region, with the smallest period 3 appearing
at $\lambda=1+\sqrt{8}=3.828427$. By the Sarkovskii ordering of natural numbers,
there is embedded in this fully chaotic region a backward directional arrow
that induces a return to lower periodic stability that eventually terminates
with the period doubling sequence in $3<\lambda\leq\lambda_{*}$. This spatial
$\lambda$-induced global dissipative decrease in $\lambda$ in the face of
the prevalent increase towards $\lambda>\lambda_{*}$ that can be taken to be
a result of the anti-world effects, is schematically summarized in Fig. \ref{fig: Complex-Chaos}
and is expressible as \begin{equation}
\begin{array}{c}
x\overset{\mathfrak{\mathfrak{{logistic}}}}\longrightarrow f_{\lambda}(x)\left\{ \begin{array}{r}
3<\lambda\leq\lambda_{*},\,0<\nu\leq1,\textrm{ self-organizing complex system}\\
\\\overset{\textrm{ininality}}\longrightarrow\lambda_{*}<\lambda\leq4,\,\nu=0,\,\textrm{dissipative complex system}\end{array}\right\} \begin{array}{r}
\overset{\textrm{anti-}}\longleftarrow\\
{\scriptstyle (\textrm{Sarkovskii})}\uparrow\\
\underset{{\textrm{effects}}}{\longrightarrow}\end{array}\end{array}\label{eq: Complex-Chaos}\end{equation}

\noindent Under normal circumstances dynamical equilibrium is attained, as elaborated
earlier, within the local \emph{temporal} (that is with respect to the iterates)
self-organizing component of the loop above. If however the system is \emph{spatially}
driven (by an increasing $\lambda$) into the dissipative region, the global
latent anti-world effects of its periodic stable windows acts as a deterrent
and, prompted by the Sarkovskii ordering, induces the system back to the self-organizing
region of equilibration. This condition of dynamical stasis is thus marked by
a balance of \emph{both the spatial and temporal effects,} with each interacting
synergetically with the other to generate an optimum dynamical state of stability.
Reference to Figs. \ref{fig: cmplx_2cycle}, \ref{fig: cmplx_4cycle} and \ref{fig: cmplx_8cycle}
clearly illustrates that new, distinguishing and non-trivial features of the
evolutionary dynamics occur at the $2^{N}$ unstable fixed points of $f_{\lambda}$
leading to the \emph{emerging} patterns that clearly characterize the resources
$\lambda$ available to the interaction. These figures also illustrate the \emph{self-organization}
induced by further passage of time by distributing this emergent pattern throughout
$X$ in the form of equivalence classes of these $2^{N}$ basic fixed points.
Panel (a) of Fig. \ref{fig: Complex-Chaos} magnifies these features of the
defining fixed points and their classes for $\lambda<\lambda_{*}$ to generate
the stable-unstable signature in the graphically convergent limit of $t\rightarrow\infty$,
essentially reflecting the competetive cohabitation of the matter-antimatter
components associated with these points. This in turn introduces a sense of
symmetry with respect to the input-output axes of the interaction that, as shown
in panel (c), is broken when $\lambda>\lambda_{*}$ with the boundary of the
{}``edge of chaos'' signalling this physical disruption with a discontinuity
in the value of the chanoxity index $\nu$. \sublabon{figure}%

Figure \ref{fig: Complex-Chaos}(d) which summarizes these observations, identifies
the self-organizing emergent region $3<\lambda\leq\lambda_{*}$ as the life
generating and sustaining complex domain (B) of the logistic interaction $f_{\lambda}$.
Below a value of 3, the resources of $f_{\lambda}$ are insufficient for supporting
life while above 3.5699456, too much {}``heat'' is produced for sustenence
of constructive competition between the opposing directions, with the forward
drive toward uniformity of ininality effectively destroying the containing reverse
competition. By contrast, Fig. \ref{fig chaos3.75,4} confirms \ref{fig: cmplx_2cycle},
\ref{fig: cmplx_4cycle} and \ref{fig: cmplx_8cycle} that independent reductionist
behaviour of the components of a system cannont generate chaos or complexity.
\begin{figure}[htbp]
\noindent \begin{center}\input{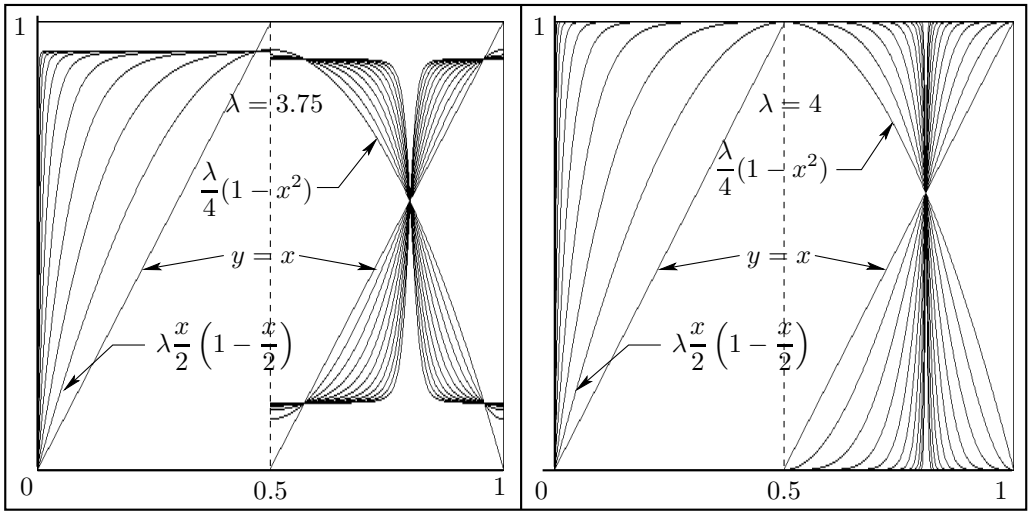}\end{center}

\caption{{\small \label{fig chaos3.75,4}Reductionism cannot generate chaos or complexity.
Together with Figs. \ref{fig: cmplx_2cycle},} \emph{\small b,} {\small and}
\emph{\small c,} {\small the present figure and this display clearly illusrates
the unique role of non-injective ill-posedness in defining chaos and complexity.
Although on its own, $\lambda(1-x^{2})/4$ can generate by splitting its unsteady
self into two steady state products, this {}``asexual'' production of {}``life''
is not nearly structurally as rich and varied as its {}``sexual'' counterpart.
A consequence of this is that unlike in its bidirectional complex logistic mode,
the unidirectional component} \emph{\small does not generate any qualitative
changes in the character of its evolutionary dynamics with a change in the value
of} {\small $\lambda$.}}
\end{figure}
\sublaboff{figure}

\section{\label{sec: Conclusions.}Conclusions: The Mechanics of Thermodynamics}

\noindent In this paper we have presented a new approach to the nonlinear dynamics
of evolutionary processes based on the mathematical framework and structure
of multifunctional graphical convergence introduced in \citet{Sengupta2003}.
The basic point we make here is that the \emph{macroscopic} dynamics of evolutionary
systems is in general governed by strongly nonlinear, non-differential laws
rather than by the Newtonian Hamilton's linear differential equations of motion
\begin{equation}
\frac{d\mathbf{x}_{i}}{dt}=\frac{\partial H(\mathbf{x})}{\partial\mathbf{p}_{i}},\quad\frac{d\mathbf{p}_{i}}{dt}=-\frac{\partial H(\mathbf{x})}{\partial\mathbf{x}_{i}},\qquad-\infty<t<\infty\label{eq: Hamilton's}\end{equation}
for the position $\mathbf{x}_{i}(t)$ and momentum $\mathbf{p}_{i}(t)$ of the
$N$ particles of an isolated (classical) system in its phase space of microstates
$\mathbf{x}(t)=(\mathbf{x}_{i}(t),\mathbf{p}_{i}(t))_{i=1}^{N}$, translated
to Liouville Equation for the macroscopic system. As is well known, Hamiltonian
dynamics leads directly to the microscopic-macroscopic paradoxes of Loschmidt's
time-reversal invariance of Eq. (\ref{eq: Hamilton's}) according to which all
forward processes of mechanical system evolving according to this law must necessarily
allow a time-reversal that would require, for example, that the Boltzmann $H$-function
decreases with time just as it increases, and Zarmelo's Poincare recurrence
paradox which postulates that almost all initial states of isolated bounded
mechanical system must recur in future, as closely as desired. One approach
--- \citet{Goldstein2004}, \citet{Price2004} --- to the resolution of these
paradoxes entail 

(1) A {}``fantastically enlarged '' phase space volume as the motivating entropy
increasing force. Thus, for example, a gas in one half of a box equilibrates
on the whole on removal of the partition so as to reach a state in which the
phase space volume is almost as large as the total phase space available to
the system under the imposed constraints, when the number of particles in the
two halves essentially become equal. In this situation, Boltzmann identifies,
\emph{for a dilute gas} of $N$ particles in a container of volume $V$ \emph{}under
weak two-body repulsive forces satisfying essentially the linearity condition
$V/N\gg b^{3}$ with $b$ the range of the force, the thermodynamic entropy
of Clausius with $S_{B}=k\ln|\Gamma(M)|$, where $\Gamma(M)$ is the region
in $6N$-dimensional Lioville phase space of the microstates belonging to the
equilibrium macrostate $M$ in question. When the system is not in equilibrium,
however, the phase space arguments imply that the relative volume of the set
of microstates corresponding to a given macrostate for which evolution leads
to a macroscopic decrease in the Boltzmann entropy \emph{typically} goes exponentially
to zero as the number of atoms in the system increases. Hence for a macroscopic
system {}``the fraction of microstates for which the evolution leads to macrostates
with larger Boltzmann entropy is so close to one that such behaviour is exactly
what should be seen to always happen'', \citet{Lebowitz1999}.

(2) The statistical techniques implicit in the foregoing interpretation of \emph{macroscopic
irreversibility in the context of microscopic reversibility of Newtonin mechanics}
rely fundamentally on the conservation of Lioville measures of sets in phase
space under evolution. This means that if a state $M(t)$ evolves as $M(t_{1})\overset{t_{1}<t_{2}}=M(t_{2})$
such that the evolved phase space $\Gamma_{t_{2}}(M(t_{1}))$ of $M(t_{1})$
is necessarily contained in $\Gamma(M(t_{2}))$ by the arguments in (1), then
the preservation of measures requires that $\Gamma_{t_{2}}(M(t_{1}))\leq\Gamma(M(t_{2}))$,
thereby verifying the increase of $S_{B}$. Identifying the macrostate of a
system with our interaction image $f(x)$ of a microstate $x$ in {}``phase
space'' $\mathscr{D}(f)$ that generates the equivalence class $\left[x\right]$
of microstates, invariance of phase space volume can be interpreted to be a
direct consequence of the \emph{linearity assumption of the Boltzmann interaction
for dilute} \emph{gases} that is also inherent in his \emph{stosszahlansatz}
assumption of molecular chaos that neglects all correlations between the particles. 

(3) Various other arguments like cosmological big bang and the relevance of
initial conditions preferring the forward direction to the reverse are invoked
to argue a justification for macroscopic irreversibility, that in the ultimate
analysis is a {}``consequence of the great disparity between microscopic and
macroscopic scales, together with the fact (or very reasonable assumption) that
what we observe in nature is typical behaviour, corresponding to typical initial
conditions'', \citet{Goldstein2004}. 

In comparison the multifunctional graphical convergence techniques, founded
on difference rather than differential equations, adapted here avoids much of
the paradoxical problems of calculus-based Hamiltonian mechanics, and suggests
an alternate specifically nonlinear dynamical framework for the dissipative
dynamical evolution of Nature that support self-organization, adaption, and
emergence in complex systems. The significant contribution of the difference
equations is that evolution at any time depends explicitly on its immediate
predecessor --- and thereby on all its predecessors --- leading to non-reductionism,
self-emergence, and complexity. 

\bigskip{}
\noindent \textbf{\large Acknowledgements}. Ii is my pleasure and privilege
to acknowledge a deep sense of gratitude to the participants of the international
workshop \emph{Mathematics and Physics of Complex and Nonlinear Systems} that
was held at IIT Kanpur, March 14-26 2004, for having elevated the proccedings
to great heights through their dedicated multiple lectures and the consequent
discussions during this 2-week period. The present paper has benefitted immensely
from the resulting collective and complex interaction that helped catalyze my
evolving ideas to the present form. 

{\small \bibliographystyle{/mnt/win/MikTex/texmf/bibtex/bst/harvard/jmps}
\bibliography{/home/osegu/LyxDocs/osegu}
}
\end{document}